\documentclass[preprint]{revtex4}
\usepackage{graphicx}
\usepackage{hyperref}
\usepackage{natbib}
\usepackage{hypernat}
\usepackage{amsmath}
\usepackage{enumerate}

\newcommand{\point}[1]{\ensuremath{{\rm #1}}}
\newcommand{\subP}{_\point{P}}
\newcommand{\Eqref}[1]{\eqref{eq:#1}}
\newcommand{\Eq}[1]{Eq.~\Eqref{#1}}
\newcommand{\Fig}[1]{Fig.~\ref{fig:#1}}
\newcommand{\Sec}[2][Section]{#1~\ref{sec:#2}}
\newcommand{\Ssec}[1]{\Sec[Subsection]{#1}}
\newcommand{\App}[1]{\Sec[Appendix]{#1}}
\newcommand{\ie}{{\em i.e.,\ }}
\newcommand{\eg}{{\em e.g.,\ }}

\begin{document}

\title{The fundamental role of the retarded potential
in the electrodynamics of superluminal sources}

\author{Houshang Ardavan}
\affiliation{University of Cambridge}
\author{Arzhang Ardavan}
\affiliation{University of Oxford}
\author{John Singleton}
\email{jsingle@lanl.gov}
\author{Joseph Fasel}
\author{Andrea Schmidt}
\affiliation{Los Alamos National Laboratory}

\begin{abstract}
We calculate the gradient
of the radiation field generated by a polarization current
with a superluminally rotating distribution pattern
and show that the absolute value of this gradient {\em increases}
as $R^{7/2}$
with distance $R$,
within the sharply focused subbeams
that constitute the overall radiation beam
from such a source.
In addition to supporting the earlier finding
that the azimuthal and polar widths of these subbeams become narrower
(as $R^{-3}$ and $R^{-1}$, respectively),
with distance from the source,
this result implies that the boundary contribution
to the solution of the wave equation governing the radiation field
does not always vanish in the limit
where the boundary tends to infinity
(as is commonly assumed in textbooks and the published literature).
There is a fundamental difference
between the classical expression for the retarded potential
and the corresponding retarded solution
of the wave equation that governs the electromagnetic field:
while the boundary contribution
to the retarded solution for the potential
can always be rendered equal to zero
by means of a gauge transformation
that preserves the Lorenz condition,
the boundary contribution
to the retarded solution of the wave equation for the field
may be neglected only if it diminishes with distance
faster than the contribution of the source density in the far zone.
In the case of a rotating superluminal source, however,
the boundary term in the retarded solution for the field
is by a factor of the order of $R^{1/2}$ {\em larger}
than the source term of this solution,
in the limit where the boundary tends to infinity.
This result is consistent
with the prediction of the retarded potential
that the radiation field generated by a rotating superluminal source
decays as $R^{-1/2}$,
instead of $R^{-1}$,
and explains why an argument
based on the solution of the wave equation governing the field
in which the boundary term is neglected
(such as that presented by J.\ H.\ Hannay)
misses the nonspherical decay of the field.
Given that the distribution of the radiation field
of an accelerated superluminal source in the far zone
is not known {\em a priori},
to be prescribed as a boundary condition,
our analysis establishes
that the only way one can calculate
the free-space radiation field of such sources
is via the retarded solution for the potential.
Finally,
we discuss the applicability of these findings
to pulsar observational data:
the more distant a pulsar,
the narrower and brighter its giant pulses should be.
\end{abstract}

\preprint{LA-UR-07-5835}

\maketitle

\section{Introduction\label{sec:intro}}
Moving sources of electromagnetic radiation
whose speeds exceed the speed of light {\em in vacuo}
have already been generated in the laboratory
\cite{BessarabAV:FasEsi,ArdavanA:Exponr,BessarabAV:Expser,BolotovskiiBM:Radssv}.
These sources arise from separation of charges:
their superluminally moving distribution patterns
are created by the coordinated motion
of aggregates of subluminally moving particles.
A polarization current density is, however,
on the same footing as the current density of free charges
in the Amp\'ere-Maxwell equation,
so that the propagating distribution patterns
of such polarization currents radiate,
as would any other moving sources of the electromagnetic field
\cite{BolotovskiiBM:VaveaD,GinzburgVL:vaveaa,BolotovskiiBM:Radbcm,ArdavanH:Genfnd,ArdavanH:Speapc}.

We have already shown,
by means of an analysis
based on the classical expression for the retarded potential
[\Eq{6} below],
that the radiation field
of a superluminally rotating extended source
at a given observation point $\point{P}$
arises almost exclusively
from those of its volume elements that approach $\point{P}$,
along the radiation direction,
with the speed of light and zero acceleration
at the retarded time
\cite{ArdavanH:Genfnd,ArdavanH:Speapc,ArdavanH:Morph}.
These elements
comprise a filamentary part of the source
whose radial and azimuthal widths become narrower
(as $\delta r\sim{R\subP}^{-2}$ and $\delta\varphi\sim{R\subP}^{-3}$,
respectively),
the larger the distance $R\subP$ of the observer from the source,
and whose length is of the order of the length scale $l_z$
of the source parallel to the axis of rotation
\cite{ArdavanH:Morph}.
($r$, $\varphi$, and $z$
are the cylindrical polar coordinates of source points.)

Once a source travels faster than its emitted waves,
it can make more than one retarded contribution
to the field that is observed at any given instant.
This multivaluedness of the retarded time
\cite{BolotovskiiBM:Radbcm,ArdavanH:Genfnd,ArdavanH:Speapc,ArdavanH:Morph}
means that the wave fronts
emitted by each of the contributing volume elements of the source
possess an envelope,
which in this case consists
of a two-sheeted, tubelike surface
whose sheets meet tangentially
along a spiraling cusp curve
(see Figs.~1 and 4 of Ref.~\cite{ArdavanH:Morph}).
For moderate superluminal speeds,
the field inside the envelope
receives contributions from three distinct values of the retarded time,
while the field outside the envelope
is influenced by only a single instant of emission time.
Coherent superposition of the emitted waves on the envelope
(where two of the contributing retarded times coalesce)
and on its cusp
(where all three of the contributing retarded times coalesce)
results in
not only a spatial,
but also a temporal focusing of the waves:
the contributions from emission
over an extended period of retarded time
reach an observer who is located on the cusp
during a significantly shorter period of observation time.

The field of each contributing volume element of the source
is strongest, therefore,
on the cusp of the envelope of wave fronts that it emits.
The bundle of cusps
generated by the collection of contributing source elements
(\ie by the filamentary part of the source
that approaches the observer with the speed of light and zero acceleration)
constitute a radiation subbeam
whose widths in the polar and azimuthal directions
are of the order of $\delta\theta\subP\sim{R\subP}^{-1}$
and  $\delta\varphi\subP\sim{R\subP}^{-3}$, respectively
\cite{ArdavanH:Morph}.
($R\subP$, $\varphi\subP$ and $\theta\subP$
are the spherical polar coordinates of the observation point $\point{P}$.)
The overall radiation beam generated by the source
consists of a
(necessarily incoherent
\cite{note:incoherent})
superposition of such subbeams,
a beam whose azimuthal width
is the same as the azimuthal extent of the source
and whose polar width
$\arccos[c/(r_<\omega)]\le|\theta\subP-\pi/2|\le\arccos[c/(r_>\omega)]$
is determined by the radial extent
$1<{\hat r}_<\le{\hat r}\le{\hat r}_>$
of the superluminal part of the source
\cite{ArdavanH:Speapc,ArdavanH:Morph}.
($c$ is the speed of light {\em in vacuo},
$\omega$ is the angular frequency of rotation of the source,
and ${\hat r}\equiv r\omega/c$.)

Since the cusps only represent the loci of points
at which the emitted spherical waves interfere constructively
(\ie represent wave packets
that are constantly dispersed and reconstructed out of other waves),
the subbeams generated by a superluminal source
need not be subject to diffraction
as are conventional radiation beams.
Nevertheless,
they have a decreasing angular width only in the polar direction.
Their azimuthal width $\delta\varphi\subP$
decreases as ${R\subP}^{-3}$ with distance
because they receive contributions
from an azimuthal extent $\delta\varphi$ of the source
that likewise shrinks as ${R\subP}^{-3}$.
They would have had a constant azimuthal width
had the azimuthal extent of the contributing part of the source
been independent of $R\subP$.
On the other hand,
the solid angle occupied by the cusps
has a thickness $\delta z\subP$
in the direction parallel to the rotation axis
that remains of the order of the height $l_z$
of the source distribution
at all distances
(see Fig.~2 of Ref.~\cite{ArdavanH:Morph}).
Consequently,
the polar width $\delta\theta\subP$
of the particular subbeam that goes through the observation point
decreases as ${R\subP}^{-1}$,
instead of being independent of $R\subP$
\cite{ArdavanH:Morph}.

Because it is of a constant linear width,
parallel to the rotation axis,
an individual subbeam subtends an area of the order of $R\subP$,
rather than ${R\subP}^2$.
In order that the flux of energy
remain the same across all cross sections of the subbeam,
therefore,
it is essential
that the Poynting vector associated with this radiation
correspondingly decay more slowly
than that of a conventional, spherically decaying radiation:
as ${R\subP}^{-1}$,
rather than ${R\subP}^{-2}$,
within the bundle of cusps
that emanate from the constituent volume elements of the source
and extend into the far zone.
This result,
which also follows
from the superposition of the Li\'enard-Wiechert fields
of the constituent volume elements
of a rotating superluminal source
\cite{ArdavanH:Speapc,ArdavanH:Morph},
has now been demonstrated experimentally
\cite{ArdavanA:Exponr}.

The narrowing of the individual subbeams with distance
suggests that the absolute value
of the gradient of the radiation field described here
should {\em increase} with distance,
in contrast to that of a conventional, diffracting radiation beam
that decreases with distance.
This is illustrated by a simple example.
Imagine a rotating radiation beam
with the amplitude
\begin{equation*}
A(R\subP, \varphi\subP, t\subP)=
A_0{{\hat R}\subP}^{1/2}\exp[-({{\hat R}\subP}^3{{\hat\varphi}\subP})^2],
\end{equation*}
where ${\hat R}\subP$ stands for the scaled distance $R\subP\omega/c$,
${\hat\varphi}\subP\equiv\varphi\subP-\omega t\subP$
is the azimuthal angle in the rotating frame,
$t\subP$ is the observation time,
and $A_0$ is a constant.
This beam would be observed
as a Gaussian pulse that has an azimuthal width
of the order of ${{\hat R}\subP}^{-3}$
and carries a constant flux of energy,
\begin{equation*}
\int A^2{R\subP}^2\sin\theta\subP{\rm d}\theta\subP{\rm d}\varphi\subP=
(2\pi)^{1/2}(c/\omega)^2{A_0}^2,
\end{equation*}
across any large sphere of radius $R\subP$.
The gradient of the amplitude of this pulse,
\begin{equation*}
\partial A/\partial{\hat\varphi}\subP=
-2A_0{{\hat R}\subP}^{7/2}({{\hat R}\subP}^3{{\hat\varphi}\subP})
\exp[-({{\hat R}\subP}^3{{\hat\varphi}\subP})^2],
\end{equation*}
increases in magnitude with distance as ${R\subP}^{7/2}$
at the edges of the pulse.

In this paper,
we derive the azimuthal (equivalently, temporal) gradient
($\partial/\partial{\hat\varphi}\subP$)
of the radiation field
that is generated by a physically viable, rotating superluminal source
directly from the retarded potential,
and show that the absolute value of this gradient
does increase as ${{\hat R}\subP}^{7/2}$
within each subbeam.
The spiky structure
of the angular distribution of the emission
from an accelerated superluminal source
therefore follows
not only from the geometry of the emitted cusps
(geometrical optics)
that was considered in Ref.~\cite{ArdavanH:Morph},
but also from the calculation of the field distribution
(physical optics)
that is presented here.
This result corroborates the earlier finding
that the overall radiation beam
consists of an incoherent superposition of sharply peaked subbeams
that become narrower with distance from the source
\cite{note:incoherent}.

There is, however, another, more significant implication.
The retarded solution
to the wave equation that governs the electromagnetic potential
in the Lorenz gauge
[\Eq{2} below]
generally entails three terms:
an integral over the retarded value of the electric current density,
an integral over the boundary values of the potential and its gradient,
and an integral over the initial values of the potential
and its time derivative
[see \Eq{3} below].
For a localized distribution of electric current,
the integral over the retarded value of the source density
is of the order of ${{\hat R}\subP}^{-1}$
in the far zone.
If evaluated for a potential
that is of this order of magnitude in the far zone
(\ie decays as ${{\hat R}\subP}^{-1}$),
the integral over the boundary in this solution
would also be of the order of ${{\hat R}\subP}^{-1}$
in the limit where the boundary tends to infinity.
However,
even potentials that satisfy the Lorenz condition
are arbitrary to within a solution of the homogeneous wave equation,
so that one can always use the gauge freedom
in the choice of potential
to set this boundary term identically equal to zero.

In the case of the corresponding retarded solution of the wave equation
for the electromagnetic field
[\Eq{7} below],
on the other hand,
one no longer has the freedom
offered by a gauge transformation
to render the boundary term equal to zero.
Nor does this term always decay faster than the source term,
so that it could be neglected for a boundary that tends to infinity,
as is commonly assumed in textbooks
(\eg page 246 of \cite{JacksonJD:Classical})
and the published literature
\cite{HannayJH:Bouffr,HannayJH:ComIGf,HannayJH:ComMhd,HannayJH:Speapc}.
The boundary contribution
to the retarded solution of the wave equation governing the field
entails a surface integral
over the boundary values of both the field and its gradient
[see \Eq{8} below].
In the superluminal regime,
where the gradient of the field increases
as ${{\hat R}\subP}^{7/2}$
over a solid angle that decreases as ${R\subP}^{-4}$,
this boundary contribution
turns out to be of the order of ${{\hat R}\subP}^{-1/2}$
(see \Sec{eval}).
Not only is this not negligible
relative to the contribution from the source term,
which decays as ${{\hat R}\subP}^{-1}$
\cite{HannayJH:Bouffr,HannayJH:ComIGf,HannayJH:ComMhd,HannayJH:Speapc},
but the boundary term constitutes the dominant contribution
toward the value of the radiation field in this case.

Thus,
if one ignores the boundary term
in the retarded solution of the wave equation governing the field
(as is done by Hannay
\cite{HannayJH:Bouffr,HannayJH:ComIGf,HannayJH:ComMhd,HannayJH:Speapc}),
one would obtain a result,
in the superluminal regime,
that contradicts what is obtained
by calculating the field via the retarded potential
\cite{ArdavanH:Genfnd,ArdavanH:Speapc,ArdavanH:Morph}.
However,
the contradiction stems solely
from having ignored a term
in the solution to the wave equation
that is by a factor of the order of ${{\hat R}\subP}^{1/2}$ greater
than the term that is normally kept in this solution.
The contradiction disappears
once the neglected term is taken into account:
the solutions to both the wave equation that governs the potential
and the wave equation that governs the field
predict that the field
of a rotating superluminal source
decays as ${{\hat R}\subP}^{-1/2}$
as $R\subP$ tends to infinity.

From a physical point of view, however,
what one obtains by including the boundary term
in the retarded solution to the wave equation
that governs the field
is merely a mathematical identity;
it is not a solution
that could be used to calculate the field
arising from a given source distribution in free space.
Unless its boundary term
happens to be negligibly smaller than its source term,
a condition that cannot be known {\em a priori},
the solution in question would require
that one prescribe the field in the radiation zone
(\ie what one is seeking)
as a boundary condition.
The role played by the classical expression
for the retarded potential in radiation theory
is clearly much more fundamental
than that played by the corresponding retarded solution
of the wave equation governing the field.
The only way to calculate the free-space radiation field
of an accelerated superluminal source
is to calculate the retarded potential
and differentiate the resulting expression
to find the field (see also \cite{ArdavanH:Speapc1}).
We must again emphasize
that this is an important contrast with subluminal sources.

This paper is organized as follows:
\Sec{bterm} presents the retarded solutions
to the initial-boundary value problems
for the wave equations that govern the potential and the field.
We provide a detailed mathematical derivation
of the gradient of the radiation field
that is generated by a rotating superluminal source
in \Sec{grad},
with a brief account of the required background material in \Ssec{back},
the formulation of the problem in \Ssec{form},
the derivation of an integral representation
of the gradient of the Green's function in \Ssec{azgrad},
the regularization of the integral
over the radial extent of the source
in \Ssec{reg}
(and \App{Hadamard}),
a description of contours of steepest descent in \Ssec{cont},
and the asymptotic evaluation of the gradient of the radiation field
in \Ssec{asymp}.
\Sec{eval} evaluates the boundary term
in the retarded solution to the wave equation governing the field,
and we conclude in \Sec{conc}.

\section{Boundary term in the solution to the wave equation\label{sec:bterm}}
In the Lorenz gauge,
the electromagnetic fields
\begin{equation}
{\bf E}=-\nabla\subP A^0-\partial{\bf A}/\partial(c t\subP),
\qquad{\bf B}=\nabla\subP\times{\bf A},
\label{eq:1}
\end{equation}
are given by a four-potential $A^\mu$
that satisfies the wave equation
\begin{equation}
{\bf\nabla}^2A^\mu-{1\over c^2}{\partial^2A^\mu\over\partial t^2}=
-{4\pi\over c}j^\mu,\qquad\mu=0,\cdots, 3,
\label{eq:2}
\end{equation}
where $A^0/c$ and $j^0/c$ are the electric potential and the charge density,
and $A^\mu$ and $j^\mu$ for $\mu=1,2,3$ are the components
of the magnetic potential ${\bf A}$ and the current density ${\bf j}$
in a Cartesian coordinate system
\cite{JacksonJD:Classical}.
The solution to the initial-boundary value problem for \Eq{2}
is given by
\begin{equation}\begin{split}
A^\mu({\bf x}\subP,t\subP)=
&{1\over c}\int_0^{t\subP}{\rm d}t\int_V{\rm d}^3x\,j^\mu G
+{1\over4\pi}\int_0^{t\subP}{\rm d}t
\int_\Sigma{\rm d}{\bf S}\cdot(G\nabla A^\mu-A^\mu\nabla G)\\
&-{1\over 4\pi c^2}\int_V{\rm d}^3x
\Big(A^\mu{\partial G\over\partial t}-G{\partial A^\mu\over\partial t}\Big)_{t=0},
\end{split}\label{eq:3}
\end{equation}
in which $G$ is the Green's function
and $\Sigma$ is the surface enclosing the volume $V$
(see, \eg page 893 of \cite{MorsePM:Methods1}).

The potential that arises
from a time-dependent localized source in unbounded space
decays as ${R\subP}^{-1}$ when $R\subP\gg1$,
so that for an arbitrary free-space potential
the second term in \Eq{3}
would be of the same order of magnitude
($\sim{R\subP}^{-1}$)
as the first term
in the limit that the boundary $\Sigma$ tends to infinity.
However,
even potentials that satisfy the Lorenz condition
${\bf\nabla\cdot A}+c^{-2}\partial A^0/\partial t=0$
are arbitrary
to within a solution of the homogeneous wave equation:
the gauge transformation
\begin{equation}
{\bf A}\to{\bf A}+\nabla\Lambda,
\qquad A^0\to A^0-\partial\Lambda/\partial t
\label{eq:4}
\end{equation}
preserves the Lorenz condition
if $\nabla^2\Lambda-c^{-2}\partial^2\Lambda/\partial t^2=0$
(see \cite{JacksonJD:Classical}).
One can always use this gauge freedom in the choice of the potential
to render the boundary contribution
(the second term)
in \Eq{3}
equal to zero,
since this term, too,
satisfies the homogenous wave equation.
Under the null initial conditions
$A^\mu|_{t=0}=(\partial A^\mu/\partial t)_{t=0}=0$,
assumed in this paper,
the contribution from the third term in \Eq{3}
is identically zero.

In the absence of boundaries,
the retarded Green's function has the form
\begin{equation}
G({\bf x}, t;{\bf x}\subP, t\subP)={\delta(t\subP-t-R/c)\over R},
\label{eq:5}
\end{equation}
where $\delta$ is the Dirac delta function
and $R$ is the magnitude of the separation
${\bf R}\equiv{\bf x}\subP-{\bf x}$
between the observation point ${\bf x}\subP$
and the source point ${\bf x}$.
Irrespective of whether the radiation decays spherically or nonspherically,
therefore,
the potential $A^\mu$ due to a localized source distribution,
which is switched on at $t=0$ in an unbounded space,
can be calculated from the first term in \Eq{3}:
\begin{equation}
A^\mu({\bf x}\subP,t\subP)=
c^{-1}\int{\rm d}^3 x{\rm d}t\, j^\mu({\bf x},t)\delta(t\subP-t-R/c)/R,
\label{eq:6}
\end{equation}
\ie from the classical expression for the retarded potential.
Whatever the Green's function for the problem may be
in the presence of boundaries,
it would approach that in \Eq{5}
in the limit where the boundaries tend to infinity,
so that one can also use this potential to calculate the field
on a boundary that lies at large distances from the source.

Next,
let us consider the wave equation that governs the magnetic field
\begin{equation}
{\bf\nabla}^2{\bf B}-{1\over c^2}{\partial^2{\bf B}\over\partial t^2}=
-{4\pi\over c}{\bf\nabla\times j}.
\label{eq:7}
\end{equation}
This may be obtained
by simply taking the curl of the wave equation for the vector potenial
[\Eq{2} for $\mu=1,2,3$].
We write the solution to the initial-boundary value problem for \Eq{7},
in analogy with \Eq{3},
as
\begin{equation}\begin{split}
B_k({\bf x}\subP,t\subP)=
&{1\over c}\int_0^{t\subP}{\rm d}t\int_V{\rm d}^3x\,({\bf\nabla\times j})_k G
+{1\over4\pi}\int_0^{t\subP}
{\rm d}t\int_\Sigma{\rm d}{\bf S}\cdot(G\nabla B_k-B_k\nabla G)\\
&-{1\over 4\pi c^2}\int_V{\rm d}^3x
\Big(
B_k{\partial G\over\partial t}-G{\partial B_k\over\partial t}
\Big)_{t=0},
\end{split}\label{eq:8}
\end{equation}
where $k=1,2,3$ designate the components of ${\bf B}$ and ${\bf\nabla\times j}$
in a Cartesian coordinate system.
Here,
we no longer have the freedom,
offered in the case of \Eq{3} by a gauge transformation,
to make the boundary term zero.

Our task in this paper
is to demonstrate that the boundary contribution in \Eq{8}
is, in fact,
by a factor of the order of ${{\hat R}\subP}^{1/2}$ larger
than the source term of this equation in the far zone
when the source is superluminal and accelerated.
For this purpose,
we need to know how the the gradient $\nabla B_k$
in the second term in \Eq{8}
decays in the far zone.
We shall calculate,
in the following section,
the field ${\bf B}$ and its gradient directly
from the classical expression for the retarded potential
[\Eq{6}],
and use the resulting expressions
to evaluate the second term in \Eq{8}
for a boundary that lies in the far zone.

\section{Gradient of the radiation field
generated by a rotating superluminal source\label{sec:grad}}
\subsection{Background: The exact expression for the radiation field\label{sec:back}}
We base our analysis
on the generic superluminal source distribution
considered in Refs.~\cite{ArdavanH:Speapc} and [10],
which has already been created in the laboratory
\cite{ArdavanA:Exponr}.
This source comprises a polarization current density
${\bf j}=\partial{\bf P}/\partial t$
for which
\begin{equation}
P_{r,\varphi,z}(r,\varphi,z,t)=
s_{r,\varphi,z}(r,z)\cos(m{\hat\varphi})\cos(\Omega t),
\qquad -\pi<{\hat\varphi}\le\pi,
\label{eq:9}
\end{equation}
with
\begin{equation}
{\hat\varphi}\equiv\varphi-\omega t,
\label{eq:10}
\end{equation}
where $P_{r,\varphi,z}$ are the components of the polarization ${\bf P}$
in a cylindrical coordinate system based on the axis of rotation,
${\bf s}(r,z)$ is an arbitrary vector
that vanishes outside a finite region of the $(r,z)$ space,
and $m$ is a positive integer.
For a fixed value of $t$,
the azimuthal dependence of the density \Eqref{8}
along each circle of radius $r$ within the source
is the same as that of a sinusoidal wave train,
of wavelength $2\pi r/m$
whose $m$ cycles fit around the circumference of the circle smoothly.
As time elapses,
this wave train both propagates around each circle
with the velocity $r\omega$
and oscillates in its amplitude
with the frequency $\Omega$.
This is a generic source:
one can construct any distribution
with a uniformly rotating pattern,
$P_{r,\varphi,z}(r,{\hat\varphi},z)$,
by the superposition over $m$
of terms of the form $s_{r,\varphi,z}(r,z,m)\cos(m{\hat\varphi})$.

To find the retarded field that follows from \Eq{6}
for the source described in \Eq{9},
we first calculated in Ref.~\cite{ArdavanH:Speapc}
the Li\'enard-Wiechert field
of a circularly moving point source
with a speed $r\omega>c$,
\ie
a generalization of the synchrotron radiation
to the superluminal regime.
We then evaluated the integral representing the retarded field
(rather than the retarded potential)
of the extended source \Eqref{8}
by superposing the fields
generated by the constituent volume elements of this source,
\ie by using the generalization of the synchrotron field
as the Green's function for the problem
(see also \cite{ArdavanH:Speapc1}).
In the superluminal regime,
this Green's function has extended singularities
that arise from the coherent superposition of the emitted waves
on the envelope of wave fronts and its cusp.

Inserting the expression
for ${\bf j}=\partial{\bf P}/\partial t$ from \Eq{9}
into \Eq{6},
and changing the variables of integration
from $({\bf x},t)=(r,\varphi,z,t)$
to $(r,\varphi,z,{\hat\varphi})$,
we found in Eq.~(20) of Ref.~\cite{ArdavanH:Speapc}
that the magnetic field ${\bf B}$ of the generated radiation
is given by
\begin{equation}
{\bf B}=-\textstyle{1\over2}{\rm i}(\omega/c)^2
\sum_{\mu=\mu_\pm}
\int_V r{\rm d}r{\rm d}{\hat\varphi}{\rm d}z\,\mu\exp(-{\rm i}\mu{\hat\varphi})
\sum_{j=1}^3{\bf u}_j\partial G_j/\partial{\hat\varphi},
\label{eq:11}
\end{equation}
where $\mu_\pm\equiv(\Omega/\omega)\pm m$,
\begin{equation}
{\bf u}_1\equiv
s_r\cos\theta\subP{\hat{\bf e}}_\parallel+s_\varphi{\hat{\bf e}}_\perp,\quad
{\bf u}_2\equiv
-s_\varphi\cos\theta\subP{\hat{\bf e}}_\parallel+s_r{\hat{\bf e}}_\perp,\quad
{\bf u}_3\equiv-s_z\sin\theta\subP{\hat{\bf e}}_\parallel,
\label{eq:12}
\end{equation}
and $G_j$ ($j=1,2,3$)
are the functions resulting
from the remaining integration with respect to $\varphi$:
\begin{equation}
\begin{bmatrix}G_1\\ G_2\\ G_3\end{bmatrix}=
\int_{\Delta\varphi} {\rm d}\varphi\,
{\delta(g-\phi)\over R}\exp({\rm i}\Omega\varphi/\omega)
\begin{bmatrix}\cos(\varphi-\varphi\subP)\\
\sin(\varphi-\varphi\subP)\\
1\end{bmatrix}.
\label{eq:13}
\end{equation}
Here
$\phi$ stands for ${\hat\varphi}-{\hat\varphi}\subP$
with ${\hat\varphi}\subP\equiv \varphi\subP-\omega t\subP$,
$R$ is
\begin{equation}
R=[(z\subP-z)^2+{r\subP}^2+r^2-2r\subP r\cos(\varphi\subP-\varphi)]^{1\over2},
\label{eq:14}
\end{equation}
the function $g$ is defined by
\begin{equation}
g\equiv\varphi-\varphi\subP+{\hat R},
\label{eq:15}
\end{equation}
with ${\hat R}\equiv R\omega/c$,
$\Delta\varphi$ is the interval of azimuthal angle traversed by the source,
and $V$ is the volume occupied by the source
in the $(r, {\hat\varphi}, z)$ space.
The unit vector
${\hat{\bf e}}_\parallel\equiv({\hat{\bf e}}_z\times{\hat{\bf n}})
/\vert{\hat{\bf e}}_z\times{\hat{\bf n}}\vert$
(which is parallel to the plane of rotation),
${\hat{\bf e}}_\perp\equiv{\hat{\bf n}}\times{\hat{\bf e}}_\parallel$,
and the radiation direction ${\hat{\bf n}}\equiv{\bf R}/R$
together form an orthonormal triad
(${\hat{\bf e}}_z$ is the base vector associated with the coordinate $z$).
The corresponding expression for the electric field
in the limit $R\subP\equiv\vert{\bf x}\subP\vert\to\infty$,
where ${\hat{\bf n}}\simeq{\bf x}\subP/\vert{\bf x}\subP\vert$,
is given by
${\bf E}={\hat{\bf n}}{\bf \times B}$,
as in any other radiation.

A distinctive feature of the emission from a superluminal source
is the multivaludeness of the retarded time
\cite{BolotovskiiBM:Radbcm,ArdavanH:Genfnd,ArdavanH:Speapc,ArdavanH:Morph}.
At any given observation time,
at least three distinct contributions,
arising from three differing retarded times,
are made toward the value of the radiation field
by the part of the source
that lies within the following volume
of the $(r,{\hat\varphi},z)$ space:
\begin{equation}
\Delta\ge0,\qquad\phi_-\le\phi\le\phi_+,
\label{eq:16}
\end{equation}
where
\begin{equation}
\Delta=({{\hat r}\subP}^2-1)({\hat r}^2-1)-({\hat z}-{\hat z}\subP)^2,
\label{eq:17}
\end{equation}
\begin{equation}
\phi_\pm=
2\pi-\arccos[(1\mp\Delta^{1\over2})/({\hat r}{\hat r}\subP)]+{\hat R}_\pm,
\label{eq:18}
\end{equation}
and
\begin{equation}
{\hat R}_\pm=
[({\hat z}-{\hat z}\subP)^2+{\hat r}^2+{{\hat r}\subP}^2-2(1\mp\Delta^{1\over2})]^{1\over2}.
\label{eq:19}
\end{equation}
This volume is bounded by a two-sheeted surface,
the so-called {\em bifurcation surface},
whose two sheets $\phi=\phi_\pm(r,z)$
meet tangentially along a cusp
(see Figs.~3 and 4 of Ref.~\cite{ArdavanH:Speapc}).
The strongest contributions
are made by the source elements
that lie close to the cusp curve $\Delta=0$,
$\phi=\phi_\pm|_{\Delta=0}$,
where the two sheets of the bifurcation surface meet tangentially.
For ${\hat R}\subP\gg1$,
the filamentary locus of these contributing source elements
is essentially parallel to the rotation axis
and has exceedingly narrow radial and azimuthal widths,
of the orders of ${{\hat R}\subP}^{-2}$ and ${{\hat R}\subP}^{-3}$,
respectively
(see Fig.~2 of Ref.~\cite{ArdavanH:Morph})

The asymptotic values of the Green's functions $G_j$
close to the cusp curve of the bifurcation surface
(where $\Delta\ll1$)
are given by
\begin{subequations}\label{eq:20}
\begin{equation}
G_j=\begin{cases}
{G_j}^{\rm in}&|\chi|<1\\
{G_j}^{\rm out}&|\chi|>1,
\end{cases}
\label{eq:20a}
\end{equation}
with
\begin{equation}
{G_j}^{\rm in}\simeq2{c_1}^{-2}(1-\chi^2)^{-{1\over2}}
[p_j\cos(\textstyle{1\over3}\arcsin\chi)
-c_1q_j\sin(\textstyle{2\over3}\arcsin\chi)],
\label{eq:20b}
\end{equation}
and
\begin{equation}
{G_j}^{\rm out}\simeq{c_1}^{-2}(\chi^2-1)^{-{1\over2}}
[p_j\sinh(\textstyle{1\over3}{\rm arccosh}|\chi|)
+c_1q_j{\rm sgn}(\chi)\sinh(\textstyle{2\over3}{\rm arccosh}|\chi|)],
\label{eq:20c}
\end{equation}
\end{subequations}
where
\begin{equation}
\chi\equiv3(\phi-c_2)/(2{c_1}^3),
\label{eq:21}
\end{equation}
with
\begin{equation}
c_1\equiv(\textstyle{3\over4})^{1\over3}(\phi_+-\phi_-)^{1\over3},\qquad
c_2\equiv\textstyle{1\over2}(\phi_++\phi_-),
\label{eq:22}
\end{equation}
and the symbol $\simeq$ denotes asymptotic approximation.
The derivation of these asymptotic values,
together with the exact expressions
for the coefficients $p_j(r,z)$ and $q_j(r,z)$
may be found in the Appendices of Refs.~\cite{ArdavanH:Genfnd} and [9].
Here, we only need the following limiting values of these coefficients
for ${\hat R}\subP\gg1$:
\begin{equation}
p_1\simeq
2^{1\over3}(\omega/c){{\hat R}\subP}^{-2}\exp({\rm i}\Omega\varphi_c/\omega),
\label{eq:23}
\end{equation}
\begin{equation}
p_2\simeq-{\hat R}\subP p_1,\quad p_3\simeq-p_2,
\label{eq:24}
\end{equation}
and
\begin{equation}
q_1\simeq
2^{2\over3}(\omega/c){{\hat R}\subP}^{-1}\exp({\rm i}\Omega\varphi_c/\omega),
\label{eq:25}
\end{equation}
\begin{equation}
q_2\simeq-q_3\simeq-{\rm i}(\Omega/\omega)q_1,
\label{eq:26}
\end{equation}
where $\varphi_c\simeq\varphi\subP+3\pi/2$ in this limit.
Note that, in these expressions,
$G_j^{\rm in,out}$
represent the different forms assumed by the Green's functions $G_j$
inside and outside the bifurcation surface,
\ie for $\phi$ inside and outside the interval $(\phi_-,\phi_+)$
respectively
(see Fig.~6 of Ref.~\cite{ArdavanH:Speapc}).

The above results show
that as a source point $(r,{\hat\varphi}, z)$
in the vicinity of the cusp curve
$\Delta=0,\phi=\phi_\pm|_{\Delta=0}$,
approaches the bifurcation surface from inside,
\ie as $\chi\to1-$ or $\chi\to-1+$,
${G_j}^{\rm in}$ and hence $G_j$ diverge.
However,
as a source point approaches
one of the sheets of the bifurcation surface
from outside,
$G_j$ tends to a finite limit:
\begin{equation}
{G_j}^{\rm out}\big|_{\phi= \phi_\pm}
={G_j}^{\rm out}\big|_{\chi=\pm1}\simeq(p_j\pm2c_1q_j)/(3{c_1}^2);
\label{eq:27}
\end{equation}
for,
the numerator of ${G_j}^{\rm out}$
is also zero when $|\chi|=1$.
The Green's function $G_j$ is singular,
in other words,
only on the inner side of the bifurcation surface
(see Fig.~6 of Ref.~\cite{ArdavanH:Speapc}).

\subsection{Formulation of the problem\label{sec:form}}
It turns out
that none of the componenets of the gradient of ${\bf B}$
can be evaluated for the source distribution \Eqref{9}
without a lengthy calculation.
However,
we shall see in \Sec{eval}
that the radial component of $\nabla B_k$
is of the same order of magnitude in the far zone
as the azimuthal (or equivalently, temporal) component
$\partial B_k/\partial{\hat\varphi}\subP$
of the gradient of $B_k(r\subP,{\hat\varphi}\subP,z\subP)$.
Since this component of the field gradient
is both algebraically simpler to calculate
and more directly related to the observeable characteristics
of the generated subbeams
(\Sec{intro}),
it will be the only component
that we shall here evaluate in detail.
The relationship between the far-field values
of this and the other components of the field gradient
is not difficult to establish
(\Sec{eval}).

The component $\partial{\bf B}/\partial{\hat\varphi}\subP$
of the gradient of ${\bf B}$
may be calculated by differentiating the right-hand side of \Eq{11}
under the integral sign
and using the fact that
$\partial G/\partial{\hat\varphi}\subP=-\partial G/\partial{\hat\varphi}$.
It follows from an argument identical to that given in Ref.~\cite{ArdavanH:Speapc}
(in connection with calculating ${\bf B}$ itself)
that the contribution
$(\partial{\bf B}/\partial{\hat\varphi}\subP)_{\Delta\ge0}$
arising from the source elements in $\Delta\ge0$
toward the value of $\partial{\bf B}/\partial{\hat\varphi}\subP$
can be written as
\begin{equation}
(\partial{\bf B}/\partial{\hat\varphi}\subP)_{\Delta\ge0}=
(\partial{\bf B}/\partial{\hat\varphi}\subP)^{\rm in}
+(\partial{\bf B}/\partial{\hat\varphi}\subP)^{\rm out}
\label{eq:28}
\end{equation}
with
\begin{subequations}\label{eq:29}
\begin{equation}
(\partial{\bf B}/\partial{\hat\varphi}\subP)^{\rm in,out}=
\textstyle{1\over2}{\rm i}(\omega/c)^2\sum_{j=1}^3
\int_{\Delta\ge0}r\,{\rm d}r\,{\rm d}z\,{\bf u}_j{L_j}^{\rm in,out},
\label{eq:29a}
\end{equation}
where
\begin{equation}
{L_j}^{\rm in}=
\sum_{\mu=\mu_\pm}\int_{\phi_-}^{\phi_+}{\rm d}\phi\,
\mu\exp(-{\rm i}\mu{\hat\varphi})
(\partial^2 G_j/\partial{\hat\varphi}^2)^{\rm in},
\label{eq:29b}
\end{equation}
and
\begin{equation}
{L_j}^{\rm out}=
\sum_{\mu=\mu_\pm}
\Big(
\int_{-\pi-{\hat\varphi}\subP}^{\phi_-}+\int_{\phi_+}^{\pi-{\hat\varphi}\subP}
\Big)
{\rm d}\phi\,
\mu\exp(-{\rm i}\mu{\hat\varphi})
(\partial^2 G_j/\partial{\hat\varphi}^2)^{\rm out}.
\label{eq:29c}
\end{equation}
\end{subequations}
Once it is integrated by parts,
the integral in \Eq{29b} in turn splits into three terms:
\begin{equation}\begin{split}
{L_j}^{\rm in}=
&\sum_{\mu=\mu_\pm}\Big\{
\mu\exp(-{\rm i}\mu{\hat\varphi})
\Big[(\partial G_j/\partial{\hat\varphi})^{\rm in}
+{\rm i}\mu {G_j}^{\rm in}\Big]
\Big|_{\phi_-}^{\phi_+}\\
&-\mu^3\int_{\phi_-}^{\phi_+}{\rm d}\phi\,
\exp(-{\rm i}\mu{\hat\varphi}){G_j}^{\rm in}\Big\},
\end{split}\label{eq:30}
\end{equation}
of which the first two (integrated) terms are divergent
[see \Eq{20b}].
Hadamard's finite part of ${L_j}^{\rm in}$
and hence of $(\partial{\bf B}/\partial{\hat\varphi}\subP)^{\rm in}$,
here designated by the prefix ${\cal F}$,
is obtained by discarding
this divergent contribution toward the value of ${L_j}^{\rm in}$
(see Refs.~\cite{ArdavanH:Speapc} and \cite{HoskinsRF:DeltaFn7}:
\begin{equation}
{\cal F}\big\{{L_j}^{\rm in}\big\}=
-\mu^3\sum_{\mu=\mu_\pm}\int_{\phi_-}^{\phi_+}
{\rm d}\phi\,\exp(-{\rm i}\mu{\hat\varphi}){G_j}^{\rm in}.
\label{eq:31}
\end{equation}
Note that the singularity of the kernel of this integral,
\ie the singularity of ${G_j}^{\rm in}$,
is like that of $|{\hat\varphi}_\pm-{\hat\varphi}|^{-{1\over2}}$
and so is integrable.

The boundary contributions from $\phi=\phi_\pm$
that result from the integration of the right-hand side of \Eq{29c} by parts
are well-defined automatically:
\begin{equation}\begin{split}
{L_j}^{\rm out}=
&-\sum_{\mu=\mu_\pm}\Big\{
\mu\exp(-{\rm i}\mu{\hat\varphi})
\Big[(\partial{G_j}/\partial{\hat\varphi})^{\rm out}
+{\rm i}\mu{G_j}^{\rm out}\Big]\Big|_{\phi_-}^{\phi_+}\\
&+\Big(\int_{-\pi-{\hat\varphi}\subP}^{\phi_-}
+\int_{\phi_+}^{\pi-{\hat\varphi}\subP}\Big)
{\rm d}\phi{\mu}^3\exp(-{\rm i}\mu{\hat\varphi}){G_j}^{\rm out}\Big\},
\end{split}\label{eq:32}
\end{equation}
since $(\partial{G_j}/\partial{\hat\varphi})^{\rm out}$
(like ${G_j}^{\rm out}$)
tends to a finite limit
as the bifurcation surface is approached from outside
(see \Ssec{azgrad}).
In deriving \Eq{32},
we have made use of the fact that
$(\partial{G_j}/\partial{\hat\varphi})^{\rm out}
|_{\phi=\pi-{\hat\varphi}\subP}$
equals
$(\partial{G_j}/\partial{\hat\varphi})^{\rm out}
|_{\phi=-\pi-{\hat\varphi}\subP}$
when $\phi_\pm\ne\pm\pi-{\hat\varphi}\subP$.
The integral representing ${L_j}^{\rm out}$,
in other words,
is finite by itself and needs no regularization.

If we now insert ${\cal F}\{{L_j}^{\rm in}\}$
and ${L_j}^{\rm out}$ from
Eqs.~\Eqref{31} and \Eqref{32}
in \Eq{29a}
and combine $(\partial{\bf B}/\partial{\hat\varphi}\subP)^{\rm in}$
and $(\partial{\bf B}/\partial{\hat\varphi}\subP)^{\rm out}$,
we arrive at an expression
for the Hadamard finite part
of $(\partial{\bf B}/\partial{\hat\varphi}\subP)_{\Delta\ge0}$
which entails both a volume and a surface integral:
\begin{equation}
{\cal F}\{(\partial{\bf B}/\partial{\hat\varphi}\subP)_{\Delta\ge0}\}=
(\partial{\bf B}/\partial{\hat\varphi}\subP)^{\rm s}
+(\partial{\bf B}/\partial{\hat\varphi}\subP)^{\rm ns}.
\label{eq:33}
\end{equation}
The volume integral
\begin{equation}\begin{split}
(\partial{\bf B}/\partial{\hat\varphi}\subP)^{\rm s}=
&-\textstyle{1\over2}{\rm i}(\omega/c)^2
\sum_{\mu=\mu_\pm}
\mu^3
\int_{\Delta\ge0}r\,{\rm d}r\,{\rm d}z\,
\int_{-\pi}^\pi {\rm d}{\hat\varphi}\,\exp(-{\rm i}\mu{\hat\varphi})\\
&\times\sum_{j=i}^3{\bf u}_jG_j
\end{split}\label{eq:34}
\end{equation}
has the same form
as the familiar integral representation
of the field of a subluminal source
\cite{JacksonJD:Classical}
and decays spherically
(as ${R\subP}^{-1}$ for ${\hat R}\subP\gg1$).

The surface integral
\begin{equation}
(\partial{\bf B}/\partial{\hat\varphi}\subP)^{\rm ns}\equiv
-\textstyle{1\over2}{\rm i}(\omega/c)^2\sum_{j=1}^3
\int_{\Delta\ge0}r\,{\rm d}r\,{\rm d}z\,{\bf u}_j{L_j}^{\rm edge}
\label{eq:35}
\end{equation}
stems from the boundary contribution
\begin{equation}
{L_j}^{\rm edge}\equiv
\sum_{\mu=\mu_\pm}\mu\exp(-{\rm i}\mu{\hat\varphi})
\Big[(\partial{G_j}/\partial{\hat\varphi})^{\rm out}
+{\rm i}\mu{G_j}^{\rm out}\Big]\Big|_{\phi_-}^{\phi_+}
\label{eq:36}
\end{equation}
in \Eq{32}.
It is this contribution
that turns out to increase,
rather than decay,
in the limit $R\subP\to\infty$.
To see this,
we need to know the values
of $(\partial{G_j}/\partial{\hat\varphi})^{\rm out}$
at $\phi=\phi_\pm$,
in addition to those of ${G_j}^{\rm out}|_{\phi=\phi_\pm}$
which are given in \Eq{27}.

\subsection{Azimuthal (or temporal) gradient of the Green's function\label{sec:azgrad}}
The Green's function \Eqref{13}
depends on ${\hat\varphi}$ only through the variable $\phi$
which appears in the argument of the Dirac delta function,
so that the differentiation of \Eq{13}
with respect to ${\hat\varphi}$
simply yields
\begin{equation}
\partial G_j/\partial{\hat\varphi}=
-\int_{\Delta\varphi}{\rm d}\varphi\, h_j(\varphi)\delta^\prime(g-\phi),
\label{eq:37}
\end{equation}
where $\delta^\prime$
stands for the derivative of the delta function
with respect to its argument,
and
\begin{equation}
\begin{bmatrix}h_1\\ h_2\\ h_3\end{bmatrix}=
{\exp({\rm i}\Omega\varphi/\omega)\over R}
\begin{bmatrix}\cos(\varphi-\varphi\subP)\\
\sin(\varphi-\varphi\subP)\\
1\end{bmatrix}.
\label{eq:38}
\end{equation}
Integrating the right-hand side of \Eq{37} by parts,
we obtain
\begin{equation}\begin{split}
{\partial G_j\over\partial{\hat\varphi}}
&=-\int_{\Delta\varphi}{\rm d}\varphi\,
{h_j(\varphi)\over\partial g/\partial\varphi}
{{\rm d}\over{\rm d}\varphi}\delta(g-\phi)\\
&=\int_{\Delta\varphi}{\rm d}\varphi\,
{{\rm d}\over{\rm d}\varphi}
\Big[{h_j(\varphi)\over\partial g/\partial\varphi}\Big]
\delta(g-\phi),
\end{split}\label{eq:39}
\end{equation}
when the source trajectory intersects
the bifurcation surface of the observation point
(\ie the argument of the delta function vanishes
within $\Delta\varphi$).
A uniform asymptotic approximation to this integral,
for small $\Delta$,
can be found by the method of Chester {\em et al.}
in the time domain
\cite{ChesterC:Extstd,BurridgeR:Asyeir}.

Where it is analytic
(\ie for all ${\bf x}\neq{\bf x}\subP$),
the function $g(\varphi)$
transforms to the cubic function
\begin{equation}
g(\varphi)=\textstyle{1\over3}\nu^3-{c_1}^2\nu+c_2,
\label{eq:40}
\end{equation}
where $\nu$ is a new variable of integration replacing $\varphi$
and the coefficients $c_1$ and $c_2$
[defined in \Eq{22}]
are such that the values
of the two functions on opposite sides of \Eq{40}
coincide at their extrema.
Insertion of \Eq{40}
and its derivative,
\begin{equation}
{\partial g\over\partial\varphi}=
{\nu^2-{c_1}^2\over{\rm d}\varphi/{\rm d}\nu}
\label{eq:41}
\end{equation}
in \Eq{39}
results in
\begin{subequations}\label{eq:42}
\begin{equation}
{\partial G_j\over\partial{\hat\varphi}}=\int_{\Delta\nu}{\rm d}\nu\,
\Big[-{F_j\over(\nu^2-{c_1}^2)^2}+{{F^\prime}_j\over\nu^2-{c_1}^2}\Big]
\delta\big(\textstyle{1\over3}\nu^3-{c_1}^2\nu+c_2-\phi\big),
\label{eq:42a}
\end{equation}
where
\begin{equation}
F_j\equiv
\Big({{\rm d}\varphi\over{\rm d}\nu}\Big)^3
{\partial^2g\over\partial\varphi^2}h_j,
\label{eq:42b}
\end{equation}
\begin{equation}
{F^\prime}_j\equiv
\Big({{\rm d}\varphi\over{\rm d}\nu}\Big)^2
{\partial h_j\over\partial\varphi},
\label{eq:42c}
\end{equation}
\end{subequations}
and $\Delta\nu$ is the image of $\Delta\varphi$
under transformation \Eqref{40}.

As in the evaluation of $G_j$
in Refs.~\cite{ArdavanH:Genfnd} and [9],
the leading term
in the asymptotic expansion of the integral \Eqref{42a}
for small $c_1$,
which corresponds to small $\Delta$
[see \Eq{53} below],
can now be obtained
by replacing the functions $F$ and $F^\prime$ in its integrand
with $P_j+Q_j\nu$ and ${P^\prime}_j+{Q^\prime}_j\nu$,
respectively,
and extending its range $\Delta\nu$
to $(-\infty,\infty)$:
\begin{subequations}\label{eq:43}
\begin{equation}
{\partial G_j\over\partial{\hat\varphi}}\simeq\int_{-\infty}^\infty{\rm d}\nu\,
\Big[-{P_j+Q_j\nu\over(\nu^2-{c_1}^2)^2}
+{{P^\prime}_j+{Q^\prime}_j\nu\over\nu^2-{c_1}^2}\Big]
\delta\big(\textstyle{1\over3}\nu^3-{c_1}^2\nu+c_2-\phi\big),
\label{eq:43a}
\end{equation}
where
\begin{equation}
P_j=\textstyle{1\over2}(F_j|_{\nu=c_1}+F_j|_{\nu=-c_1}),
\label{eq:43b}
\end{equation}
\begin{equation}
Q_j=\textstyle{1\over2}{c_1}^{-1}(F_j|_{\nu=c_1}-F_j|_{\nu=-c_1}),
\label{eq:43c}
\end{equation}
\begin{equation}
{P^\prime}_j=
\textstyle{1\over2}({F^\prime}_j|_{\nu=c_1}+{F^\prime}_j|_{\nu=-c_1}),
\label{eq:43d}
\end{equation}
and
\begin{equation}
{Q^\prime}_j=
\textstyle{1\over2}{c_1}^{-1}({F^\prime}_j|_{\nu=c_1}
-{F^\prime}_j|_{\nu=-c_1}).
\label{eq:43e}
\end{equation}
\end{subequations}
Note that the extrema
\begin{equation}
\varphi_\pm=
2\pi-\arccos[(1\mp\Delta^{1\over2})/({\hat r}{\hat r}\subP)]
\label{eq:44}
\end{equation}
of the function $g(\varphi)$
transform into $\nu=\mp c_1$,
respectively.

The derivatives $d\varphi/d\nu|_{\nu=\pm c_1}$
that appear in the definitions
of the coefficients $(P_j,Q_j,{P^\prime}_j, {Q^\prime}_j)$
are indeterminate.
Their values must be found
by repeated differentiation of Eqs.~\Eqref{15} and (40)
with respect to $\nu$:
\begin{subequations}\label{eq:45}
\begin{equation}
(dg/d\varphi)(d\varphi/d\nu)=\nu^2-{c_1}^2,
\label{eq:45a}
\end{equation}
\begin{equation}
(d^2g/d\varphi^2)(d\varphi/d\nu)^2+(dg/d\varphi)(d^2\varphi/d\nu^2)=2\nu,
\label{eq:45b}
\end{equation}
\end{subequations}
etc.,
and the evaluation of the resulting relations at $\nu=\pm c_1$.
This procedure,
which amounts to applying l'H\^opital's rule,
yields
\begin{equation}
d\varphi/d\nu|_{\nu=\pm c_1}=
(2c_1{\hat R}_{\mp})^{1\over2}/\Delta^{1\over4}.
\label{eq:46}
\end{equation}
Using $\partial^2g/\partial\varphi^2|_{\varphi_\pm}
=\mp\Delta^{1/2}/{\hat R}_\pm$
and \Eq{46},
we find from \Eq{42b} that
\begin{equation}
F_j\big|_{\nu=\pm c_1}=
\pm 2c_1 f_j\big|_{\nu=\pm c_1},
\label{eq:47}
\end{equation}
in which $f_j=({\rm d}\varphi/{\rm d}\nu)h_j$
are the functions earlier encountered
in the evaluation of $G_j$
in Refs.~\cite{ArdavanH:Genfnd} and [9].
Hence, $P_j=2{c_1}^2q_j$ and $Q_j=2p_j$,
where $p_j$ and $q_j$ are precisely the same
as the coefficients in Eqs.~\Eqref{20}
that are approximated in Eqs.~\Eqref{23}--\Eqref{26}
(see Ref.~\cite{ArdavanH:Speapc}).

We now need to evaluate $\partial G_j/\partial{\hat\varphi}$
only outside the bifurcation surface,
\ie for $|\chi|>1$
[see Eqs.~\Eqref{21} and \Eqref{36}].
In this region,
the argument of the $\delta$ function in \Eq{43a}
has a single zero at
\begin{equation}
\nu=\nu^*=2c_1{\rm sgn}(\chi)
\cosh(\textstyle{1\over3}{\rm arccosh}|\chi|),\qquad
|\chi|\ge1,
\label{eq:48}
\end{equation}
(see Appendix~A of Ref.~\cite{ArdavanH:Genfnd}).
Outside the bifurcation surface, therefore,
\Eq{43a} yields
\begin{equation}
\Big({\partial {G_j}\over\partial{\hat\varphi}}\Big)^{\rm out}\simeq
{1\over|\nu^2-{c_1}^2|}
\Big[-{{P_j+Q_j\nu}\over(\nu^2-{c_1}^2)^2}
+{{{P^\prime}_j+{Q^\prime}_j\nu}\over\nu^2-{c_1}^2}\Big]
\Big|_{\nu^*}.
\label{eq:49}
\end{equation}
Keeping only the first term in this expresssion,
which is dominant when $c_1\ll1$,
we obtain
\begin{equation}
\Big({\partial {G_j}\over\partial{\hat\varphi}}\Big)^{\rm out}\simeq
-{2\sinh^3(\textstyle{1\over3}{\rm arccosh}|\chi|)
\over{c_1}^5(\chi^2-1)^{3/2}}
\big[c_1q_j+2p_j{\rm sgn}(\chi)
\cosh(\textstyle{1\over3}{\rm arccosh}|\chi|)\big],
\label{eq:50}
\end{equation}
in which $p_j(r,z)$ and $q_j(r,z)$ assume
the values given in Eqs.~\Eqref{23}--\Eqref{26} when ${\hat R}\subP\gg1$.

Evaluation of the right-hand side of \Eq{50} at $\phi=\phi_\pm$
now yields the following term
that appears in the expression for ${L_j}^{\rm edge}$:
\begin{equation}\begin{split}
\exp(-{\rm i}\mu{\hat\varphi})
(\partial {G_j}/\partial{\hat\varphi})^{\rm out}
\big|_{\phi_-}^{\phi_+}\simeq
&\textstyle-({2\over3})^3{c_1}^{-5}\exp[-{\rm i}\mu({\hat\varphi}\subP+c_2)]\\
&\times\big[p_j\cos(\textstyle{2\over3}\mu{c_1}^3)
-\textstyle{1\over2}{\rm i}c_1 q_j\sin(\textstyle{2\over3}\mu{c_1}^3)\big].
\end{split}\label{eq:51}
\end{equation}
The asymptotic expansions of
${G_j}^{\rm out}$ and $(\partial G_j/\partial{\hat\varphi})^{\rm out}$
given in Eqs.~\Eqref{20c} and \Eqref{50}
are for small $c_1$.
To be consistent,
we must likewise replace
the expression that is found by inserting ${G_j}^{\rm out}$
and $(\partial G_j/\partial{\hat\varphi})^{\rm out}$
in \Eq{36}
with the leading term in its expansion in powers of $c_1$.
The result is
\begin{equation}
{L_j}^{\rm edge}\simeq
-2^{1/3}(\textstyle{2\over3})^3{R\subP}^{-1}{\bar p}_j{c_1}^{-5}
\exp({\rm i}\Omega\varphi_c/\omega)
\sum_{\mu=\mu_\pm}\mu\exp[-{\rm i}\mu({\hat\varphi}\subP+\phi_-)],
\label{eq:52}
\end{equation}
in which ${\bar p}_j\equiv({{\hat R}\subP}^{-1}\quad-1\quad+1)$
and $\varphi_c\simeq\varphi\subP+3\pi/2$.

The far-field value of $c_1$
close to the cusp curve of the bifurcation surface
(where $\Delta=0$)
is given by
\begin{equation}
c_1\simeq2^{-{1\over3}}{{\hat R}\subP}^{-1}\Delta^{1\over2}
\label{eq:53}
\end{equation}
[see \Eq{22}].
Inserting \Eq{52} in \Eq{35}
and using \Eq{53},
we finally arrive at
\begin{equation}\begin{split}
(\partial{\bf B}/\partial{\hat\varphi}\subP)^{\rm ns}\simeq
&2(\textstyle{2\over3})^3{{\hat R}\subP}^4
\exp[{\rm i}(\Omega\varphi_c/\omega-\pi/2)]
\sum_{\mu=\mu_\pm}\mu\exp(-{\rm i}\mu{\hat\varphi}\subP)\\
&\times\sum_{j=1}^3{\bar p}_j
\int_{\Delta\ge0}{\hat r}\,{\rm d}{\hat r}\,{\rm d}{\hat z}\,
\Delta^{-{5\over2}}{\bf u}_j\exp(-{\rm i}\mu\phi_-).
\end{split}\label{eq:54}
\end{equation}
As in Ref.~\cite{ArdavanH:Morph},
the integral over ${\hat r}$ in this expression
may be evaluated by contour integration.
Since the singularity of its integrand at $\Delta=0$
is not integrable, however,
the contour integral that passes through this singularity
needs to be in addition regularized.

\subsection{Regularization of the integral over the radial extent of the source\label{sec:reg}}
The kernel of the integral in \Eq{54}
has the same phase but a different amplitude
compared to that of the integral encountered in Eq.~(19) of Ref.~\cite{ArdavanH:Morph}.
Hence, the asymptotic evaluation of integral \Eqref{54}
entails the same techniques as those used before,
but the regularization of this integral
requires an extension of the procedure followed in Ref.~\cite{ArdavanH:Morph}.

The function $\phi_-({\hat r},{\hat z})$
in the phase of the integrand in \Eq{54}
is stationary as a function of ${\hat r}$ at
\begin{equation}
{\hat r}={\hat r}_C({\hat z})\equiv
\{\textstyle{1\over2}({{\hat r}\subP}^2+1)
-[\textstyle{1\over4}({{\hat r}\subP}^2-1)^2
-({\hat z}-{\hat z}\subP)^2]^{1/2}\}^{1/2}.
\label{eq:55}
\end{equation}
When the observer is located in the far zone,
this one-dimensional locus of stationary points coincides with the locus,
\begin{equation}
{\hat r}={\hat r}_S\equiv[1+({\hat z}-{\hat z}\subP)^2/({{\hat r}\subP}^2-1)]^{1/2},
\label{eq:56}
\end{equation}
of source points
that approach the observer with the speed of light and zero acceleration
at the retarded time,
\ie with the projection $\Delta=0$
of the cusp curve of the bifurcation surface
onto the $(r,z)$ plane
(see Fig.~4 of Ref.~\cite{ArdavanH:Morph}).
For ${\hat R}\subP\gg1$,
the separation ${\hat r}_C-{\hat r}_S$
vanishes as ${{\hat R}\subP}^{-2}$
[see \Eq{69} below]
and both ${\hat r}_C$ and ${\hat r}_S$
assume the value $\csc\theta\subP$.

It follows from \Eq{18}
that at the stationary point ${\hat r}={\hat r}_C$,
\begin{equation}
\phi_-|_{{\hat r}={\hat r}_C}\equiv
\phi_C={\hat R}_C+\varphi_C-\varphi\subP,
\label{eq:57}
\end{equation}
$\partial\phi_-/\partial{\hat r}|_{{\hat r}={\hat r}_C}=0$,
and
\begin{equation}
\partial^2\phi_-/\partial{\hat r}^2\big|_{{\hat r}={\hat r}_C}\equiv
a=-{{\hat R}_C}^{-1}[({{\hat r}\subP}^2-1)({{\hat r}_C}^2-1)^{-1}-2],
\label{eq:58}
\end{equation}
where
\begin{equation}
\varphi_C=\varphi\subP+2\pi-\arccos({\hat r}_C/{\hat r}\subP)
\quad{\rm and}\quad
{\hat R}_C={\hat r}_C({{\hat r}\subP}^2-{{\hat r}_C}^2)^{1\over2}.
\label{eq:59}
\end{equation}
For observation points of interest to us
(the observation points located outside the plane of rotation,
$\theta\subP\neq\pi/2$, in the far zone, ${\hat R}\subP\gg1$),
the parameter $a$ in \Eq{58}
has a value whose magnitude increases with increasing ${\hat R}\subP$:
\begin{equation}
a\simeq-{\hat R}\subP\sin^4\theta\subP\sec^2\theta\subP.
\label{eq:60}
\end{equation}
In other words,
the phase function $\phi_-$ is more sharply peaked at its maximum,
the farther the observation point is from the source.
This property of the phase function $\phi_-$
distinguishes the asymptotic analysis
that will be presented in this section
from those commonly encountered in radiation theory.
What turns out to play the role of a large parameter
in this asymptotic expansion
is distance (${\hat R}\subP$),
not frequency ($\mu_\pm\omega$).

The first step in the asymptotic analysis
of the integral in \Eq{54}
is to introduce a change of variable
$\xi=\xi({\hat r},{\hat z})$
that replaces the original phase $\phi_-$ of the integrand
by as simple a polynomial as possible
\cite{BleisteinN:Asei}.
This transformation should be one-to-one
and should preserve the number and nature
of the stationary points of the phase.
Since $\phi_-$ has a single isolated stationary point
at ${\hat r}={\hat r}_C({\hat z})$,
it can be cast into a canonical form
by means of the following transformation:
\begin{equation}
\phi_-({\hat r}, {\hat z})
=\phi_C({\hat z})+\textstyle{1\over2}a({\hat z})\xi^2,
\label{eq:61}
\end{equation}
in which $a$ is the coefficient given in Eqs.~\Eqref{58} and \Eqref{60}.

The integral in \Eq{54} can thus be written as
\begin{equation}
\int_{\Delta\ge0}{\hat r}{\rm d}{\hat r}\,{\rm d}{\hat z}\,
\Delta^{-5/2}{\bf u}_j\exp(-{\rm i}\mu\phi_-)=
\int_{\xi\ge\xi_S}{\rm d}{\hat z}{\rm d}\xi\,
F(\xi,{\hat z})\exp({\rm i}\alpha\xi^2),
\label{eq:62}
\end{equation}
in which
\begin{equation}
F(\xi,{\hat z})\equiv
{\hat r}\Delta^{-5/2}{\bf u}_j(\partial{\hat r}/\partial\xi)
\exp(-{\rm i}\mu\phi_C),
\label{eq:63}
\end{equation}
with
\begin{equation}
\partial{\hat r}/\partial\xi=
a\xi{\hat r}{\hat R}_-({\hat r}^2-1-\Delta^{1/2})^{-1},
\label{eq:64}
\end{equation}
and $\alpha\equiv-\mu a/2$.
The stationary point ${\hat r}={\hat r}_C$
and the boundary point ${\hat r}={\hat r}_S$
respectively map onto $\xi=0$ and
\begin{equation}
\xi=\xi_S\equiv-[2a^{-1}(\phi_S-\phi_C)]^{1/2},
\label{eq:65}
\end{equation}
where
\begin{equation}
\phi_S\equiv\phi_-|_{{\hat r}={\hat r}_S}
=2\pi-\arccos[1/({\hat r}_S{\hat r}\subP)]+({{\hat r}_S}^2{{\hat r}\subP}^2-1)^{1/2}.
\label{eq:66}
\end{equation}
The upper limit of integration in \Eq{62}
is determined by the image
of the support of the source density (${\bf s}$ in ${\bf u}_j$)
under the transformation \Eqref{61}.

By substituting the value of $r_{\point C}$ from \Eq{55} in \Eq{17},
we find that $\Delta^{1/2}={\hat r}^2-1$ at ${\point C}$.
Thus,
the Jacobian $\partial{\hat r}/\partial\xi$ of the above transformation
is indeterminate at $\xi=0$.
Its value at this critical point
must be found by repeated differentiation of \Eq{61} with respect to $\xi$,
\begin{equation}
(\partial\phi_-/\partial{\hat r})(\partial{\hat r}/\partial\xi)=a\xi,
\label{eq:67}
\end{equation}
\begin{equation}
(\partial^2\phi_-/\partial{\hat r}^2)(\partial{\hat r}/\partial\xi)^2
+(\partial\phi_-/\partial{\hat r})(\partial^2{\hat r}/\partial\xi^2)
=a,
\label{eq:68}
\end{equation}
and the evaluation of the resulting relation \Eqref{68}
at ${\hat r}={\hat r}_C$
with the aid of \Eq{58}.
This procedure,
which amounts to applying l'H\^opital's rule,
yields $\partial{\hat r}/\partial\xi|_{\xi=0}=1$:
a result we could have anticipated
from the coincidence of transformation \Eqref{61}
with the Taylor expansion of $\phi_-$
about ${\hat r}={\hat r}_C$
to within the leading order.
Correspondingly,
the amplitude $F(\xi)$ in \Eq{63}
has the value
${\hat r}_\point{C}({{\hat r}_\point{C}}^2-1)^{-5}{\bf u}_j|_{{\hat r}={\hat r}_\point{C}}
\exp(-{\rm i}\mu\phi_\point{C})$
at the critical point $\point{C}$.

To an observer in the far field (${\hat R}\subP\gg1$),
the phase of the integrand on the right-hand side of \Eq{62}
is rapidly oscillating,
irrespective of how low the harmonic numbers $\mu_\pm$
(\ie the radiation frequencies $\mu_\pm\omega$)
may be.
The leading contribution
to the asymptotic value of integral \Eqref{62}
from the stationary point $\xi=0$
can therefore be determined by the method of stationary phase.
However,
in the limit ${\hat R}\subP\to\infty$,
$\xi_S$ reduces to
\begin{equation}
\xi_S\simeq-3^{-1/2}\cos^4\theta\subP\csc^5\theta\subP{{\hat R}\subP}^{-2},
\label{eq:69}
\end{equation}
so that the stationary point $\xi=0$
is separated from the boundary point $\xi=\xi_S$
by an interval of the order of ${{\hat R}\subP}^{-2}$ only.
We therefore need to employ
a technique for the asymptotic analysis of integral \Eqref{62}
that is capable of handling
the contributions from both ${\hat r}_\point{C}$ and ${\hat r}_\point{S}$
simultaneously.

\subsection{Contours of steepest descent\label{sec:cont}}
The technique we shall employ for this purpose
is the method of steepest descents
\cite{BleisteinN:Asei}.
We regard the variable of integration in
\begin{equation}
J({\hat z})\equiv
\int_{\xi_S}^{\xi_>}{\rm d}\xi\, F(\xi,{\hat z})\exp({\rm i}\alpha\xi^2)
\label{eq:70}
\end{equation}
as complex,
\ie write $\xi=u+{\rm i}v$,
and invoke Cauchy's integral theorem
to deform the original path of integration
into the contours of steepest descent
that pass through each of the critical points
$\xi=\xi_S$, $\xi=0$ and $\xi=\xi_>$.
Here,
we have introduced the real variable $\xi_>({\hat z})$
to designate the image of ${\hat r}_>$ under transformation \Eqref{61},
\ie the boundary of the support of the source term ${\bf u}_j$
that appears in the amplitude $F(\xi,{\hat z})$.
We shall only treat the case
in which $\mu$ (and hence $\alpha$) is positive;
$J({\hat z})$ for negative $\mu$
can then be obtained
by taking the complex conjugate of the derived expression
and replacing $\phi_\point{C}$ with $-\phi_\point{C}$
[see \Eq{63}].

\begin{figure}[htbp]
\centering
\includegraphics[width=8.3cm]{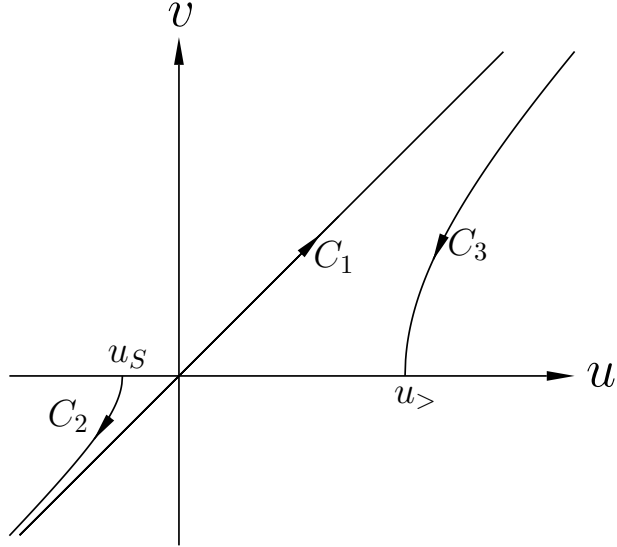}
\caption[The integration contours in the complex plane $\xi=u+{\rm i}v$]%
{The integration contours in the complex plane $\xi=u+{\rm i}v$.
The critical point $\point{C}$ lies at the origin
and $u_S$ and $u_>$ are the images under transformation \Eqref{61}
of the radial boundaries ${\hat r}={\hat r}_S({\hat z})$ and ${\hat r}={\hat r}_>({\hat z})$
of the part of the source that lies within $\Delta>0$.
The contours $C_1$, $C_2$, and $C_3$
are the paths of steepest descent
through the stationary point $\point{C}$
and through the lower and upper boundaries of the integration domain,
respectively.}
\label{fig:intCont}
\end{figure}

The path of steepest descent
through the stationary point $\point{C}$
at which $\xi=0$ is given,
according to
\begin{equation}
{\rm i}\xi^2=-2uv+{\rm i}(u^2-v^2),
\label{eq:71}
\end{equation}
by $u=v$ when $\alpha$ is positive.
If we designate this path by $C_1$
(see \Fig{intCont}),
then
\begin{equation}\begin{split}
\int_{C_1}{\rm d}\xi & F(\xi,{\hat z})\exp({\rm i}\alpha\xi^2)=
(1+{\rm i})
\int_{-\infty}^{\infty}{\rm d}v F|_{\xi=(1+{\rm i})v}\exp(-2\alpha v^2)\\
&\simeq(2\pi/\mu)^{1/2}\exp[-{\rm i}(\mu\phi_\point{C}-\pi/4)]
{\bf u}_j|_\point{C}\sin^7\theta\subP|\sec\theta\subP|^9{{\hat R}\subP}^{-1/2},
\end{split}\label{eq:72}
\end{equation}
for ${\hat R}\subP\gg1$.
Here,
we have obtained the leading term
in the asymptotic expansion of the above integral
for large ${\hat R}\subP$
by approximating $F|_{\xi=(1+{\rm i})v}$
by its value at $C$,
where $v=0$,
and using Eqs.~\Eqref{55} and \Eqref{60}
to replace ${\hat r}_\point{C}$ and $\alpha$
by their values in the far zone.
Note that the next term in this asymptotic expansion
is smaller by a factor of order ${{\hat R}\subP}^{-1/2}$
than this leading term.

The path of steepest descent
through the boundary point $\point{S}$,
at which $u\equiv u_S=\xi_S$ and $v=0$
[see Eqs.~\Eqref{65} and \Eqref{69}],
is given by $u=-(v^2+{u_S}^2)^{1/2}$,
\ie by the contour designated as $C_2$ in \Fig{intCont}.
The real part of
\begin{equation}
{\rm i}\xi^2|_{C_2}=2v(v^2+{u_S}^2)^{1/2}+{\rm i}{u_S}^2,
\label{eq:73}
\end{equation}
is a monotonic function of $v$
and so can be used as a curve parameter for contour $C_2$
in place of $v$.
If we let $2v(v^2+{u_S}^2)^{1/2}\equiv-\tau^2$,
then it follows from
\begin{equation}
\xi|_{C_2}=-({u_S}^2+{\rm i}\tau^2)^{1/2}
\label{eq:74}
\end{equation}
that
\begin{equation}\begin{split}
\int_{C_2}{\rm d}\xi& F(\xi,{\hat z})\exp({\rm i}\alpha\xi^2)=
\exp[{\rm i}(\alpha{u_S}^2-\pi/2)]\\
&\times\int_0^\infty{\rm d}\tau\,\tau({u_S}^2+{\rm i}\tau^2)^{-1/2}F
\big|_{\xi=-({u_S}^2+{\rm i}\tau^2)^{1/2}}\exp(-\alpha\tau^2).
\end{split}\label{eq:75}
\end{equation}
The function $F|_{C_2}$ in this expression
has to be determined by inverting
the following version of the original transformation \Eqref{61}:
\begin{equation}
\phi_-({\hat r},{\hat z})-\phi_S({\hat z})=\textstyle{1\over2}{\rm i}a\tau^2.
\label{eq:76}
\end{equation}
Here,
we have used Eqs.~\Eqref{65} and \Eqref{74}
to rewrite \Eq{61} in terms of $\tau$.

Since the dominant contribution
toward the asymptotic value of the above integral for ${\hat R}\subP\gg1$
comes from the vicinity of the boundary point $\point{S}$,
where $\tau=0$,
the required inversion of transformation \Eqref{76}
can be carried out
by means of a Taylor expansion of the phase function $\phi_-$
in powers of $\tau$
(see \App{Hadamard}).
We find in \App{Hadamard}
that the resulting expression for $F(\tau)|_{C_2}$
diverges at $\tau=0$ as $\tau^{-4}$.
Therefore,
as in the case of the integral over ${\hat\varphi}$ in \Eq{11},
we must regard
the divergent integral in \Eq{75}
as a generalized function
that equals its Hadamard's finite part (see, \eg Ref.~\cite{ArdavanH:Speapc1}).
The procedure for finding the Hadamard finite part of this integral,
though lengthy,
is straightforward
and results in
\begin{equation}\begin{split}
{\cal F}
\Big\{\int_{C_2}{\rm d}\xi\, F(\xi,{\hat z})\exp({\rm i}\alpha\xi^2)\Big\}=
&(35/6^4)(2\pi/\mu)^{1/2}\exp[-{\rm i}(\mu\phi_S-3\pi/4)]\\
&\times{\bf u}_j\big|_S\sin^7\theta\subP
|\sec\theta\subP|^9{{\hat R}\subP}^{-1/2}
\end{split}\label{eq:77}
\end{equation}
in the limit ${\hat R}\subP\gg1$
(see \App{Hadamard}).

The path of steepest descent
through the boundary point $\xi=\xi_>$,
at which $u=u_>$, $v=0$,
is given by $u=(v^2+u_>)^{1/2}$,
\ie by the contour designated as $C_3$ in \Fig{intCont}.
The real part of the exponent
\begin{equation}
{\rm i}\xi^2|_{C_3}=-2v(v^2+{u_>}^2)^{1/2}+{\rm i}{u_>}^2
\label{eq:78}
\end{equation}
is again a monotonic function of $v$
and so can be used to parametrize contour $C_3$ in place of $v$.
If we let $2v(v^2+{u_>}^2)^{1/2}\equiv\chi$,
then it follows from
\begin{equation}
\xi|_{C_3}=({u_>}^2+{\rm i}\chi)^{1/2}
\label{eq:79}
\end{equation}
that
\begin{equation}\begin{split}
\int_{C_3}{\rm d}\xi& F(\xi,{\hat z})\exp({\rm i}\alpha\xi^2)=
\textstyle{1\over2}\exp[{\rm i}(\alpha{u_>}^2-\pi/2)]\\
&\times\int_0^\infty{\rm d}\chi\,
({u_>}^2+{\rm i}\chi)^{-1/2}F\big|_{\xi=({u_>}^2+{\rm i}\chi)^{1/2}}\exp(-\alpha\chi).
\end{split}\label{eq:80}
\end{equation}
The asymptotic value of this integral for ${\hat R}\subP\gg1$
receives its dominant contribution from $\chi=0$.
Because the function $F|_{C_3}$ is regular,
on the other hand,
its value at $\chi=0$
can be found by simply evaluating the expression in \Eq{63}
at ${\hat r}={\hat r}_>$.
The result, for ${\hat R}\subP\to\infty$, is
\begin{equation}
F|_{C_3,\chi=0}\simeq
{{\hat r}_>}^2{{\hat R}\subP}^{-4}\sin^4\theta\subP\sec^2\theta\subP
({{\hat r}_>}^2\sin^2\theta\subP-1)^{-3}
{\bf u}_j|_{{\hat r}={\hat r}_>}\exp(-{\rm i}\mu\phi_\point{C})u_>
\label{eq:81}
\end{equation}
[see Eqs.~\Eqref{17} and \Eqref{60}].
This in conjunction with Watson's lemma therefore implies that
\begin{equation}\begin{split}
\int_{C_3}{\rm d}\xi F(\xi,{\hat z})\exp({\rm i}\alpha\xi^2)\simeq
&{{\hat r}_>}^2({{\hat r}_>}^2\sin^2\theta\subP-1)^{-3}
{\bf u}_j|_{{\hat r}={\hat r}_>}\\
&\times\exp[-{\rm i}(\phi_-|_{{\hat r}={\hat r}_>}+\pi/2)]
\mu^{-1}{{\hat R}\subP}^{-5},
\end{split}\label{eq:82}
\end{equation}
to within the leading order in ${{\hat R}\subP}^{-1}$.

\subsection{Asymptotic value of the gradient of the field for large distances\label{sec:asymp}}
The integral in \Eq{70}
equals the sum of the three contour integrals
in Eqs.~\Eqref{72}, \Eqref{77}, and \Eqref{82};
the contributions of the contours
that connect $C_1$ and $C_2$, and $C_1$ and $C_3$, at infinity
(see \Fig{intCont})
are exponentially small
compared to those of $C_1$, $C_2$, and $C_3$ themselves.
On the other hand,
the leading term in the asymptotic value
of the integral over $C_3$
decreases (with increasing ${\hat R}\subP$)
much faster than those
in the asymptotic values of the integrals over $C_1$ and $C_2$:
the integral over $C_3$
decays as ${{\hat R}\subP}^{-5}$,
while the integrals over $C_1$ and $C_2$
decay as ${{\hat R}\subP}^{-1/2}$.
According to Eqs.~\Eqref{54}, \Eqref{62}, \Eqref{72} and \Eqref{77},
the leading term
in the asymptotic expansion
of the contribution
$(\partial{\bf B}/\partial{\hat\varphi}\subP)^{\rm ns}$,
for large ${\hat R}\subP$,
is therefore given by
\begin{equation}\begin{split}
(\partial{\bf B}/\partial{\hat\varphi}\subP)^{\rm ns}\simeq&
\textstyle{1\over3^7}(35-6^4{\rm i}){{\hat R}\subP}^{7/2}\sin^7\theta\subP
|\sec\theta\subP|^9\exp\{{\rm i}[(\Omega/\omega)\varphi_\point{C}+\pi/4]\}\\
&\times\sum_{\mu=\mu_\pm}(2\pi|\mu|)^{1/2}{\rm sgn}(\mu)
\exp({\rm i}\textstyle{\pi\over4}{\rm sgn}\,\mu)\\
&\times\sum_{j=1}^3 {\bar p}_j\int_{-\infty}^\infty{\rm d}{\hat z}
\exp[-{\rm i}\mu(\phi_\point{C}+{\hat\varphi}\subP)]{\bf u}_j|_\point{C},
\end{split}\label{eq:83}
\end{equation}
in which $\mu_\pm$ can also be negative
(see the first paragraph of \Sec{cont}).

The remaining ${\hat z}$ integration
in the above expression
for $(\partial{\bf B}/\partial{\hat\varphi}\subP)^{\rm ns}$
amounts to a Fourier decomposition
of the source densities $s_{r,\varphi,z}|_\point{C}$
with respect to ${\hat z}$.
Using Eqs.~\Eqref{57}--\Eqref{59}
to replace $\phi_\point{C}$ in \Eq{83} by its far-field value
\begin{equation}
\phi_\point{C}\simeq{\hat R}\subP-{\hat z}\cos\theta\subP+3\pi/2,
\label{eq:84}
\end{equation}
and using \Eq{12} to write out ${\bf u}_j$
in terms of $s_{r,\varphi,z}$,
we find that
\begin{equation}\begin{split}
(\partial{\bf B}/\partial{\hat\varphi}\subP)^{\rm ns}\simeq
&\textstyle{1\over3^7}(35-6^4{\rm i}){{\hat R}\subP}^{7/2}
\sin^7\theta\subP|\sec\theta\subP|^9\\
&\times\exp\{{\rm i}[\pi/4+(\Omega/\omega)(\varphi\subP+3\pi/2)]\}\\
&\times\sum_{\mu=\mu_\pm}(2\pi|\mu|)^{1/2}{\rm sgn}(\mu)
\exp\{{\rm i}[\textstyle{\pi\over4}{\rm sgn}(\mu)-\mu({\hat R}\subP
+{\hat\varphi}\subP+3\pi/2)]\}\\
&\times[({\bar s}_\varphi\cos\theta\subP-{\bar s}_z\sin\theta\subP)
{\hat{\bf e}}_\parallel
-{\bar s}_r{\hat{\bf e}}_\perp],
\end{split}\label{eq:85}
\end{equation}
where ${\bar s}_{r,\varphi,z}$
stand for the following Fourier transforms of $s_{r,\varphi,z}|_\point{C}$
with respect to ${\hat z}$:
\begin{equation}
{\bar s}_{r,\varphi,z}\equiv
\int_{-\infty}^\infty{\rm d}{\hat z}\,
s_{r,\varphi,z}({\hat r},{\hat z})\big|_{{\hat r}=\csc\theta\subP}
\exp({\rm i}\mu{\hat z}\cos\theta\subP).
\label{eq:86}
\end{equation}
Being the contribution
from the source elements that approach the observer
with the speed of light and zero acceleration at the retarded time,
this expression is valid
only at those polar angles $\theta\subP$
within the interval
$\arccos(1/{\hat r}_<)\le|\theta\subP-\pi/2|\le\arccos(1/{\hat r}_>)$
for which $s_{r,\varphi,z}|_{{\hat r}=\csc\theta\subP}$ is nonzero,
\ie at those observation points
(outside the plane of rotation)
the cusp curve of whose bifurcation surface intersects the source distribution.
At these polar angles,
the above expression
for $(\partial{\bf B}/\partial{\hat\varphi}\subP)^{\rm ns}$
constitutes the dominant contribution
toward the gradient $\partial{\bf B}/\partial{\hat\varphi}\subP$
of the magnetic field of the radiation
(see \Ssec{form}).

\section{Evaluation of the boundary term
in the retarded solution to the wave equation governing the field\label{sec:eval}}
Let the boundary $\Sigma$ in the second term of \Eq{8}
be a large sphere enclosing the source.
The element ${\rm d}{\bf S}$ of area for this boundary
then has the form
$\rho^2\sin\theta{\rm d}\theta{\rm d}\varphi{\hat{\bf e}}_\rho$,
where $(\rho,\varphi,\theta)$
are the spherical polar coordinates in the space of source points,
\ie are related
to the cylindrical polar coordinates $(r,\varphi,z)$ we have been using
by
\begin{equation}
\rho\equiv(r^2+z^2)^{1/2},\qquad\theta\equiv\arctan(r/z),
\label{eq:87}
\end{equation}
and ${\hat{\bf e}}_\rho$ is a unit vector
in the direction of increasing $\rho$.
Inserting this in the integrand of the boundary contribution in \Eq{8},
we obtain
\begin{equation}
{\bf B}_{\rm boundary}=
\rho^2\int {\rm d}t\int_\Sigma{\rm d}\varphi\,{\rm d}\theta\,
\sin\theta\big(G\partial{\bf B}/\partial R\subP\big|_{R\subP=\rho}
-{\bf B}\partial G/\partial\rho\big),
\label{eq:88}
\end{equation}
since
$({\hat{\bf e}}_\rho\cdot\nabla){\bf B}=\partial{\bf B}/\partial\rho
=\partial{\bf B}/\partial R\subP|_{R\subP=\rho}$.
We will be identifying the magnetic  field ${\bf B}$ and its gradient
on the boundary $\Sigma$
with those of the radiation field that arises from source \Eqref{9}.
These terms,
which act as densities of two-dimensional sources in \Eq{88},
both have rigidly rotating distribution patterns,
\ie are functions of $t$
in the combination ${\hat\varphi}=\varphi-\omega t$ only
[see \Eq{85}].

Once the free-space Green's function \Eqref{5} is inserted in \Eq{88},
we can therefore cast it in the same form as \Eq{11}
by changing the integration variable $t$ to ${\hat\varphi}$,
resulting in
\begin{equation}
{\bf B}_{\rm boundary}=
{\hat\rho}^2\int_\Sigma{\rm d}{\hat\varphi} \,{\rm d}\theta\,
\sin\theta\big(G_b\partial{\bf B}/\partial{\hat R}\subP\big|_{R\subP=\rho}
-{\bf B}\partial G_b/\partial{\hat\rho}\big),
\label{eq:89}
\end{equation}
with
\begin{equation}
G_b\equiv\int{\rm d}\varphi{\hat R}^{-1}\delta(g-\phi),
\label{eq:90}
\end{equation}
where ${\hat\rho}=\rho\omega/c$, and $g$ and $\phi$
are the same functions as those appearing in Eqs.~\Eqref{13}--\Eqref{15}.
Equation \Eqref{90} implies that
\begin{equation}\begin{split}
G_b
&=\sum_{\varphi=\varphi_j}
{1\over{\hat R}|\partial g/\partial\varphi|}\\
&=\sum_{\varphi=\varphi_j}
|{\hat R}+{\hat\rho}{\hat R}\subP\sin\theta\sin\theta\subP
\sin(\varphi_j-\varphi\subP)|^{-1},
\end{split}\label{eq:91}
\end{equation}
where $\varphi_j$ are the solutions
of the transcendental equation $g(\varphi)=\phi$
[see \Eq{15}].
For ${\hat\rho}\gg1$,
the number of retarded positions $\varphi_j$
of the rapidly rotating distribution patterns
of ${\bf B}|_\Sigma$
and $\partial{\bf B}/\partial{\hat R}\subP|_\Sigma$
that contribute toward the value of $G_b$
can be appreciably larger than three
(see Ref.~\cite{BolotovskiiBM:Radbcm}).

The expression in \Eq{11} for the magnetic field ${\bf B}$
depends on $R\subP$ through $\partial G_j/\partial{\hat\varphi}$ only,
so that
\begin{equation}\begin{split}
\partial{\bf B}/\partial{\hat R}\subP=
&-\textstyle{1\over2}{\rm i}(\omega/c)^2\sum_{\mu=\mu_\pm}
\int_V r{\rm d}r{\rm d}{\hat\varphi}{\rm d}z\,
\mu\exp(-{\rm i}\mu{\hat\varphi})\\
&\times\sum_{j=1}^3{\bf u}_j
\partial^2 G_j/\partial{\hat R}\subP\partial{\hat\varphi}.
\end{split}\label{eq:92}
\end{equation}
with
\begin{equation}
\partial^2 G_j/\partial{\hat R}\subP\partial{\hat\varphi}=
-(\omega/c)
\int_{\Delta\varphi}{\rm d}\varphi
h_j(\varphi){\hat R}^{-1}[\delta^{\prime\prime}(g-\phi)
-{\hat R}^{-1}\delta^{\prime}(g-\phi)]\partial{\hat R}/\partial{\hat R}\subP,
\label{eq:93}
\end{equation}
where $h_j$ are the functions defined in \Eq{38},
and a prime denotes differentiation of the delta function
with respect to its argument
[see \Eq{37}].

It follows from a comparison
of the calculations described in Subsections~\ref{sec:back} and \ref{sec:azgrad}
that the contribution of the second term on the right-hand side of \Eq{93}
toward the value of $\partial{\bf B}/\partial {\hat R}\subP$
is by a factor of the order of ${{\hat R}\subP}^{-4}$
smaller than that of the first term.
Ignoring this small term,
we obtain an expression
for $-\partial^2 G_j/\partial{\hat R}\subP\partial{\hat\varphi}$
that differs from the expression
for $\partial^2 G_j/\partial{\hat\varphi}^2$
only by the factor
\begin{equation}
\partial{\hat R}/\partial{\hat R}\subP=
{\hat R}^{-1}
[{\hat R}\subP-{\hat z}\cos\theta\subP-{\hat r}\sin\theta\subP\cos(\varphi-\varphi\subP)]
\label{eq:94}
\end{equation}
[see \Eq{14}],
which reduces to $1$ in the limit ${\hat R}\subP\gg1$.
Correspondingly,
the leading contribution,
$(\partial{\bf B}/\partial{\hat R}\subP)^{\rm ns}$,
toward the value of $\partial{\bf B}/\partial{\hat R}\subP$
is given by an expression identical to that in \Eq{85}
for $(\partial{\bf B}/\partial\varphi\subP)^{\rm ns}$,
except that it is multiplied by $-1$
(see \Sec{grad}).
The absolute value of $\partial{\bf B}/\partial{\hat R}\subP$
is therefore of the same order of magnitude
as that of $\partial^2 G_j/\partial{\hat R}\subP\partial{\hat\varphi}$,
and so increases as ${{\hat R}\subP}^{7/2}$
(see \Ssec{asymp}).

We already know
that the radiation subbeams
that are generated by the superluminal source \Eqref{9}
have the widths $\delta\theta\sim{{\hat R}\subP}^{-1}$
and $\delta{\hat\varphi}\sim{{\hat R}\subP}^{-3}$
(see Ref.~\cite{ArdavanH:Morph}).
In the limit where the values of ${\hat\rho}$ and ${\hat R}\subP$
(\ie the positions of the boundary and the observer)
tend to infinity independently of each other,
the Green's function $G_b$ reduces to
\begin{equation}
G_b\simeq
\sum_{\varphi=\varphi_j}
\big[{\hat\rho}{\hat R}\subP\sin\theta\sin\theta\subP|\sin(\varphi_j-\varphi\subP)|\big]^{-1}.
\label{eq:95}
\end{equation}
Hence,
the absolute values of $G_b$ and $\partial G_b/\partial{\hat\rho}$
diminish with distance
as $({\hat\rho}{\hat R}\subP)^{-1}$ and ${\hat\rho}^{-2}{{\hat R}\subP}^{-1}$,
respectively.
Since $|\partial{\bf B}/\partial{\hat R}\subP|$ increases as ${{\hat R}\subP}^{7/2}$
while $|{\bf B}|$ decreases as ${{\hat R}\subP}^{-1/2}$,
this means that,
of the two terms inside the parantheses in \Eq{89},
the second is negligibly smaller than the first.
Inserting the orders of magnitude of the remaining factors in \Eq{89},
in the order in which they appear in the first term of this equation,
we obtain
\begin{equation}\begin{split}
|{\bf B}_{\rm boundary}|
&\sim{\hat\rho}^2\times{\hat\rho}^{-3}\times{\hat\rho}^{-1}
\times({\hat\rho}{\hat R}\subP)^{-1}\times{\hat\rho}^{7/2}\\
&\sim{\hat\rho}^{1/2}{{\hat R}\subP}^{-1}.
\end{split}\label{eq:96}
\end{equation}
Thus,
the absolute value of the boundary contribution toward the value of the field
decays as ${{\hat R}\subP}^{-1/2}$
when the radius $\rho$ of the spherical boundary $\Sigma$
and the coordinate $R\subP$ of the observation point $P$
are both large and of the same order of magnitude.

\section{Concluding remarks\label{sec:conc}}
The unaviodably lengthy calculation
that we have presented in Sections~\ref{sec:grad} and \ref{sec:eval}
both lends support to the conclusions of Ref.~\cite{ArdavanH:Morph},
on the morphology of the radiation beam
that is generated by a polarization current
with a superluminally rotating distribution pattern,
and clarifies a fundamental issue
concerning the method of calculating the radiation field
of such a polarization current,
which has been the source of a long-standing controversy
\cite{HannayJH:Bouffr,HannayJH:ComIGf,HannayJH:ComMhd,HannayJH:Speapc,ArdavanH:Speapc1}.
This calculation establishes
\begin{enumerate}[(i)]
\item
that the absolute values
of both the radial component $\partial{\bf B}/\partial{\hat R}\subP$
and the azimuthal or temporal component $\partial{\bf B}/\partial{\hat\varphi}\subP$
of the gradient of the radiation field
that is generated by the superluminal source distribution \Eqref{9} in the far zone
are of the order of ${{\hat R}\subP}^{7/2}$
at any observation point
within the overall radiation beam arising from this source
\cite{note:gradcomp},
\item
that the angular distribution of the emitted field
contains sharply focused structures,
\ie that the overall radiation beam
is composed of an incoherent superposition of rapidly narrowing subbeams
\cite{note:incoherent},
\item
that the boundary contribution
toward the the solution of the wave equation governing the field
decays as ${{\hat R}\subP}^{-1/2}$ as the boundary tends to infinity,
\ie that the second term in \Eq{8} is by a factor of the order of ${{\hat R}\subP}^{1/2}$
greater than the first term in this equation
for a $\Sigma$ that lies in ${\hat R}\subP\gg1$,
\item
that the discrepancy between the predictions of \Eq{2} and \Eq{7}
disappears once one includes the boundary term
that is normally neglected
in solution \Eqref{8},
and
\item
that Hannay's erroneous contention
that the field of a rotating superluminal source should diminish as ${R\subP}^{-1}$,
as does a conventional radiation field
\cite{HannayJH:Bouffr,HannayJH:ComIGf,HannayJH:ComMhd,HannayJH:Speapc},
stems from his having neglected the boundary term
in the solution [\Eq{8}] to the wave equation governing the field [\Eq{7}].
\end{enumerate}

The sharply focused radiation pulses encountered in the present analysis
are in fact observed in astronomical objects
that are thought to contain superluminal sources.
The radio emission received from pulsars
is composed
(often entirely
\cite{PopovMV:GPmcrecp})
of a collection of so-called giant pulses
whose widths are as narrow as $1$ ns
\cite{HankinsTH:Nanrbs}
and whose brightness temperatures are as high as $10^{39}$ K
\cite{SoglasnovVA:GiapPB}.
Hankins {\em et al} \cite{HankinsTH:Nanrbs}
note the puzzling brightness of these pulses:
\begin{quote}
The plasma structures responsible for these emissions
must be smaller than one meter in size,
making them by far the smallest objects
ever detected and resolved outside the Solar System,
and the brightest transient radio sources in the sky.
\end{quote}

The small size of the emitting structures reflects,
in the present context,
the narrowing (as ${R\subP}^{-2}$ and ${R\subP}^{-3}$, respectively)
of the radial and azimuthal dimensions
of the filamentary part of the source that approaches the observer at $\point{P}$
with the speed of light and zero acceleration at the retarded time
\cite{ArdavanH:Morph}.
This,
together with the nonspherical decay
of the individual subbeams generated by such filaments
(as ${R\subP}^{-1/2}$ instead of ${R\subP}^{-1}$),
easily accounts for the observationally inferred values
of the brightness temperature of the giant pulses.

The azimuthal (or temporal) gradient of the intensity of these pulses
often appears infinitely sharp at either their leading or trailing edges
(see Fig.~1 of Ref.~\cite{HankinsTH:Nanrbs}).
Correspondingly,
the emission mechanism discussed in this paper
sets no upper limit on the gradient $\partial/\partial{\hat\varphi}\subP$
of the radiation field
(\ie on the sharpness of the leading or trailing edge of the pulse),
if the length scale of spatial or temporal variations of its source
are comparable with ${{\hat R}\subP}^{-3}$.
According to the superluminal model of pulsars
\cite{SchmidtA:Occopmlw}
(to which the present findings apply),
the more distant a pulsar, the narrower and brighter its giant pulses should be.

\section*{Acknowledgements}
H.\ A.\ thanks Boris Bolotovskii for his help and encouragement,
and Alexander Schekochihin and Janusz Gill
for their stimulating questions and comments.
A.\ A.\ is supported by the Royal Society.
J.\ S., J.\ F., and A.\ S.\ are supported by U.S.\ Department of Energy grant LDRD 20050540ER.

\appendix

\section{Hadamard's finite part of the divergent contribution
from the integral over the contour $C_2$\label{sec:Hadamard}}
Our first task in this Appendix
is to make the dependence of the integrand
of integral \Eqref{75} on the integration variable $\tau$ explicit.
This entails
(i) inverting \Eq{76} in the vicinity of the critical point $\tau=0$
to obtain ${\hat r}$ as a function of $\tau$
for a fixed value of ${\hat z}$,
and (ii)
expanding the function $F|_{C_2}$
that appears in the integrand of integral \Eqref{75}
in powers of $\tau$.
Next, we calculate
the Hadamard finite part of the resulting integral
(whose integrand turns out to diverge as $\tau^{-4}$ at $\tau=0$)
by following the standard procedure
used in the literature on generalized functions
\cite{HoskinsRF:DeltaFn7}.

Because it contains the factor
\begin{equation}
\Delta^{1/2}=({\hat r}^2-1)^{1/2}({\hat r}^2-{{\hat r}_S}^2)^{1/2},
\label{eq:A1}
\end{equation}
the function $\phi_-({\hat r},{\hat z})$
is not analytic at ${\hat r}={\hat r}_S$
[see Eqs.~\Eqref{17}, \Eqref{18} and \Eqref{56}].
If, however,
we eliminate ${\hat r}$ in $\phi_-$
in favour of
\begin{equation}
\eta\equiv({\hat r}^2-{{\hat r}_S}^2)^{1/2},
\label{eq:A2}
\end{equation}
then the resulting function
\begin{equation}\begin{split}
\phi_-(\eta,{\hat z})=
&\{{{\hat r}\subP}^2({{\hat r}_S}^2+\eta^2)-[1+({{\hat r}\subP}^2-1)^{1/2}\eta]^2\}^{1/2}\\
&+2\pi-\arccos\{{{\hat r}\subP}^{-1}({{\hat r}_S}^2+\eta^2)^{-1/2}
[1+({{\hat r}\subP}^2-1)^{1/2}\eta]\}
\end{split}\label{eq:A3}
\end{equation}
can be expanded into a Taylor series about $\eta=0$ to obtain
\begin{equation}\begin{split}
\phi_-=
&\phi_S+\textstyle{1\over2}{{\hat R}\subP}^{-1}\cos^2\theta\subP\eta^2
-\textstyle{1\over3}\sin^3\theta\subP\eta^3
+\textstyle{1\over8}{{\hat R}\subP}^{-3}\cos^2\theta\subP(5\sin^2\theta\subP-1)\eta^4\\
&+\textstyle{1\over5}\sin^5\theta\subP\eta^5+\cdots.
\end{split}\label{eq:A4}
\end{equation}
Here,
the coefficients in this series
are approximated by their dominant values for ${\hat R}\subP\gg1$,
and the coordinate ${\hat r}_\point{S}$ that appears in them
is replaced by its value $\csc\theta\subP$
at the radius from which the main contributions toward the field in the far zone arise
[see \Eq{56}].
Equation \Eqref{A4},
in conjunction with \Eq{76},
provides us with an analytic expression for $\tau(\eta)$
that we can invert to find $\eta$
(and hence ${\hat r}$)
as a function of $\tau$.

Repeated differentiations of \Eq{76} with respect to $\tau$
result in
\begin{subequations}\label{eq:A5}
\begin{equation}
(\partial\phi_-/\partial\eta)(\partial\eta/\partial\tau)={\rm i}a\tau,
\label{eq:A5a}
\end{equation}
\begin{equation}
(\partial\phi_-/\partial\eta)(\partial^2\eta/\partial\tau^2)
+\partial^2\phi_-/\partial\eta^2)(\partial\eta/\partial\tau)^2
={\rm i}a,
\label{eq:A5b}
\end{equation}
\end{subequations}
and so on,
which when evaluated at $\point{S}$
(where $\eta=\tau=0$),
yield $\partial\eta/\partial\tau|_S$, $\partial^2\eta/\partial\tau^2|_S$, etc.,
in terms of the known derivatives
$\partial\phi_-/\partial\eta|_S$, $\partial^2\phi_-/\partial\eta^2|_S$, etc.,
that constitute the coefficients in \Eq{A4}.
Using these derivatives of $\eta$ at $\point{S}$,
we can therefore write down the Taylor expansion of $\eta$ in powers of $\tau$:
\begin{subequations}\label{eq:A6}
\begin{equation}
{\hat\eta}=
\textstyle{1\over3}{\hat\tau}+{\hat\tau}^2+\textstyle{5\over2}{\hat\tau}^3+8{\hat\tau}^4
+\textstyle{231\over8}{\hat\tau}^5+\cdots,
\label{eq:A6a}
\end{equation}
where
\begin{equation}
{\hat\eta}\equiv\textstyle{1\over3}{\hat R}\subP\sin^3\theta\subP\sec^2\theta\subP\eta,
\label{eq:A6b}
\end{equation}
and
\begin{equation}
{\hat\tau}\equiv
\textstyle{1\over3}\exp(-{\rm i}\pi/4){{\hat R}\subP}^2\sin^5\theta\subP\sec^4\theta\subP\tau.
\label{eq:A6c}
\end{equation}
\end{subequations}
The dependence of ${\hat r}$ on $\tau$ now follows from Eqs.~\Eqref{A2} and \Eqref{A6}.

According to Eqs.~\Eqref{63}, \Eqref{64}, \Eqref{74}, \Eqref{A1} and \Eqref{A2},
the explicit form of the function $F|_{C_2}$
is given by
\begin{equation}\begin{split}
F\big|_{\xi=-({u_S}^2+{\rm i}\tau^2)^{1/2}}\simeq
&\textstyle{1\over3}{{\hat R}\subP}^{-5}\csc^5\theta\subP{\bf u}_j({\xi_S}^2+{\rm i}\tau^2)^{1/2}\\
&\times\exp[-{\rm i}(\mu\phi_\point{C}-\pi/2)]\partial(\eta^{-3})/\partial\tau\big|_{C_2}
\end{split}\label{eq:A7}
\end{equation}
for ${\hat R}\subP\gg1$.
Insertion of this expression in \Eq{70} yields
\begin{equation}\begin{split}
\int_{C_2}{\rm d}\xi F(\xi,{\hat z})\exp({\rm i}\alpha\xi^2)=
&\textstyle{1\over3}{{\hat R}\subP}^{-8}\cot^6\theta\subP\csc^5\theta\subP
\exp[-{\rm i}(\mu\phi_S+\pi/4)]\\
&\times{\bf u}_j(3{\cal I}_1-{\cal I}_2)\big|_{C_2},
\end{split}\label{eq:A8}
\end{equation}
in which
\begin{equation}
{\cal I}_1\equiv\int_0^\infty{\rm d}\tau\,\tau^{-4}\exp(-\alpha\tau^2)\psi(\tau),
\label{eq:A9}
\end{equation}
and
\begin{equation}
{\cal I}_2\equiv\int_0^\infty{\rm d}\tau\,\tau^{-3}\exp(-\alpha\tau^2)({\rm d}\psi/{\rm d}\tau),
\label{eq:A10}
\end{equation}
with $\psi\equiv({\hat\eta}/{\hat\tau})^{-3}$.
Here,
we have used the fact that $\mu\phi_\point{C}-\alpha{u_S}^2=\mu\phi_\point{S}$,
where $\phi_\point{S}$ stands for the value of $\phi_-$ at $\point{S}$.

That the integrals ${\cal I}_1$ and ${\cal I}_2$ have turned out to diverge
is a consequence of our having interchanged the orders of integration and differentiation
in \Eq{37}
[see also \Eq{13}].
The standard technique for regularizing such divergent integrals
is to treat them as generalized functions
whose physically significant values
(\ie the values that we would have found
had we not interchanged the orders of integration and differentiation)
are given by their Hadamard finite parts
\cite{HoskinsRF:DeltaFn7}.

To apply the technique to ${\cal I}_1$,
one begins by appealing to Taylor's Theorem
to represent the continuously differentiable function $\psi$ as
\begin{equation}
\psi(\tau)=
\psi(0)+\psi^\prime(0)\tau+\textstyle{1\over2}\psi^{\prime\prime}(0)\tau^2
+\textstyle{1\over3!\ }\psi^{\prime\prime\prime}(0)\tau^3
+\textstyle{1\over4!\ }\psi^{\prime\prime\prime\prime}(\kappa\tau)\tau^4,
\label{eq:A11}
\end{equation}
where $\kappa$ is a number lying between $0$ and $1$.
One then inserts \Eq{A11} in \Eq{A9}
to rewrite ${\cal I}_1$ as
\begin{equation}\begin{split}
{\cal I}_1=
&\lim_{\epsilon\to0}\big[\psi(0)\int_\epsilon^\infty{\rm d}\tau\,\tau^{-4}\exp(-\alpha\tau^2)
+\psi^\prime(0)\int_\epsilon^\infty{\rm d}\tau\,\tau^{-3}\exp(-\alpha\tau^2)\\
&+\textstyle{1\over2}\psi^{\prime\prime}(0)\int_\epsilon^\infty{\rm d}\tau\,
\tau^{-2}\exp(-\alpha\tau^2)
+\textstyle{1\over3!\ }\psi^{\prime\prime\prime}(0)
\int_\epsilon^\infty{\rm d}\tau\,\tau^{-1}\exp(-\alpha\tau^2)\\
&+\textstyle{1\over4!\ }\int_\epsilon^\infty{\rm d}\tau\,
\psi^{\prime\prime\prime\prime}(\kappa\tau)\exp(-\alpha\tau^2)\big].
\end{split}\label{eq:A12}
\end{equation}
The first four integrals inside the square brackets in this expression
can be easily evaluated as functions of $(\alpha,\epsilon)$;
\eg
\begin{equation}
\int_\epsilon^\infty{\rm d}\tau\,\tau^{-4}\exp(-\alpha\tau^2)=
\textstyle{1\over3}\epsilon^{-3}(1-2\alpha\epsilon^2)\exp(-\alpha\epsilon^2)
+\textstyle{2\over3}\pi^{1/2}\alpha^{3/2}{\rm erfc}(\alpha^{1/2}\epsilon),
\label{eq:A13}
\end{equation}
in which the error function ${\rm erfc}(\alpha^{1/2}\epsilon)$
approaches unity in the limit $\epsilon\to0$.

The remaining fifth integral on the right-hand side of \Eq{A12} equals
\begin{equation}\begin{split}
\int_\epsilon^\infty{\rm d}\tau\,\psi^{\prime\prime\prime\prime}(\kappa\tau)\exp(-\alpha\tau^2)=
&4!\,\int_\epsilon^\infty{\rm d}\tau\,
\tau^{-4}[\psi(\tau)-\psi(0)-\psi^\prime(0)\tau-\textstyle{1\over2}\psi^{\prime\prime}(0)\tau^2\\
&-\textstyle{1\over3!\ }\psi^{\prime\prime\prime}(0)\tau^3]\exp(-\alpha\tau^2)
\end{split}\label{eq:A14}
\end{equation}
by virtue of \Eq{A11}.
For $\alpha\gg1$
(\ie ${\hat R}\subP\gg1$)
and $\epsilon=0$,
the leading term in the asymptotic value of the right-hand integral in \Eq{A14}
is given by
\begin{equation}
\int_0^\infty{\rm d}\tau\,\psi^{\prime\prime\prime\prime}(\kappa\tau)\exp(-\alpha\tau^2)\simeq
\psi^{\prime\prime\prime\prime}(0)\int_0^\infty{\rm d}\tau\exp(-\alpha\tau^2)
=\textstyle{1\over2}(\pi/\alpha)^{1/2}\psi^{\prime\prime\prime\prime}(0).
\label{eq:A15}
\end{equation}
Here,
we have applied l'H\^opital's rule
to remove the indeterminacy
in the value of the kernel of the right-hand integral in \Eq{14}
at $\tau=0$.

Hadamard's finite part of the limiting version
of each of the integrals that appear inside the square brackets in \Eq{A12}
is obtained by simply discarding those terms
in its representation as a function of $(\alpha,\epsilon)$
that diverge when $\epsilon$ tends to zero;
\eg
\begin{equation}
{\cal F}\Big\{\int_0^\infty{\rm d}\tau\,\tau^{-4}\exp(-\alpha\tau^2)\Big\}=
\textstyle{2\over3}\pi^{1/2}\alpha^{3/2}
\label{eq:A16}
\end{equation}
according to \Eq{A13}.
Thus,
\Eq{A15} and the finite parts of the divergent integrals on the right-hand side of \Eq{A12}
jointly yield
\begin{equation}\begin{split}
{\cal F}\{{\cal I}_1\}\simeq
&\textstyle{2\over3}\pi^{1/2}\psi(0)\alpha^{3/2}
+\textstyle{1\over2}\psi^\prime(0)(\ln\alpha+\gamma)\alpha
-\textstyle{1\over2}\pi^{1/2}\psi^{\prime\prime}(0)\alpha^{1/2}\\
&-\textstyle{1\over12}\psi^{\prime\prime\prime}(0)(\ln\alpha+\gamma)
+\textstyle{1\over48}\pi^{1/2}\psi^{\prime\prime\prime\prime}(0)\alpha^{-1/2},\qquad
\alpha\gg1,
\end{split}\label{eq:A17}
\end{equation}
where $\gamma=0.57721$ is Euler's constant.
The same procedure,
when applied to the integral defined in \Eq{A10},
results in
\begin{equation}\begin{split}
{\cal F}\{{\cal I}_2\}\simeq
&\textstyle{1\over2}\psi^\prime(0)(\ln\alpha+\gamma)\alpha
-\pi^{1/2}\psi^{\prime\prime}(0)\alpha^{1/2}
-\textstyle{1\over4}\psi^{\prime\prime\prime}(0)(\ln\alpha+\gamma)\\
&+\textstyle{1\over12}\pi^{1/2}\psi^{\prime\prime\prime\prime}(0)\alpha^{-1/2},\qquad
\alpha\gg1.
\end{split}\label{eq:A18}
\end{equation}
The required derivatives of $\psi$ at $\tau=0$
can be read off the following expansion of $({\hat\eta}/{\hat\tau})^{-3}$:
\begin{equation}
\psi=
1-3{\hat\tau}-\textstyle{3\over2}{\hat\tau}^2-4{\hat\tau}^3-\textstyle{105\over8}{\hat\tau}^4
+\cdots,
\label{eq:A19}
\end{equation}
which follows from \Eq{A6a}
[see also \Eq{A6c}].

Evaluating the right-hand sides of Eqs.~\Eqref{A17} and \Eqref{A18}
with the aid of Eqs.~\Eqref{A6c} and \Eqref{A19},
and inserting the resulting expressions in \Eq{A8},
we finally arrive at
\begin{equation}
{\cal F}\{3{\cal I}_1-{\cal I}_2\}=
-(105/6^4)\sin^{18}\theta\subP|\sec\theta\subP|^{15}{{\hat R}\subP}^{15/2},\qquad
{\hat R}\subP\gg1,
\label{eq:A20}
\end{equation}
and hence, at \Eq{77}.

\bibliography{boundary-notes,jabosa,superluminal}

\begin{thebibliography}{27}
\expandafter\ifx\csname natexlab\endcsname\relax\def\natexlab#1{#1}\fi
\expandafter\ifx\csname bibnamefont\endcsname\relax
  \def\bibnamefont#1{#1}\fi
\expandafter\ifx\csname bibfnamefont\endcsname\relax
  \def\bibfnamefont#1{#1}\fi
\expandafter\ifx\csname citenamefont\endcsname\relax
  \def\citenamefont#1{#1}\fi
\expandafter\ifx\csname url\endcsname\relax
  \def\url#1{\texttt{#1}}\fi
\expandafter\ifx\csname urlprefix\endcsname\relax\def\urlprefix{URL }\fi
\providecommand{\bibinfo}[2]{#2}
\providecommand{\eprint}[2][]{\url{#2}}

\bibitem[{\citenamefont{Bessarab et~al.}(2004)\citenamefont{Bessarab, Gorbunov,
  Martynenko, and Prudkoy}}]{BessarabAV:FasEsi}
\bibinfo{author}{\bibfnamefont{A.~V.} \bibnamefont{Bessarab}},
  \bibinfo{author}{\bibfnamefont{A.~A.} \bibnamefont{Gorbunov}},
  \bibinfo{author}{\bibfnamefont{S.~P.} \bibnamefont{Martynenko}},
  \bibnamefont{and} \bibinfo{author}{\bibfnamefont{N.~A.}
  \bibnamefont{Prudkoy}}, \bibinfo{journal}{IEEE Trans. Plasma Sci.}
  \textbf{\bibinfo{volume}{32}}, \bibinfo{pages}{1400} (\bibinfo{year}{2004}),
  ISSN \bibinfo{issn}{0093-3813}.

\bibitem[{\citenamefont{Ardavan
  et~al.}(2004{\natexlab{a}})\citenamefont{Ardavan, Hayes, Singleton, Ardavan,
  Fopma, and Halliday}}]{ArdavanA:Exponr}
\bibinfo{author}{\bibfnamefont{A.}~\bibnamefont{Ardavan}},
  \bibinfo{author}{\bibfnamefont{W.}~\bibnamefont{Hayes}},
  \bibinfo{author}{\bibfnamefont{J.}~\bibnamefont{Singleton}},
  \bibinfo{author}{\bibfnamefont{H.}~\bibnamefont{Ardavan}},
  \bibinfo{author}{\bibfnamefont{J.}~\bibnamefont{Fopma}}, \bibnamefont{and}
  \bibinfo{author}{\bibfnamefont{D.}~\bibnamefont{Halliday}},
  \bibinfo{journal}{J. Appl. Phys.} \textbf{\bibinfo{volume}{96}},
  \bibinfo{pages}{7760} (\bibinfo{year}{2004}{\natexlab{a}}), ISSN
  \bibinfo{issn}{0021-8979}, \bibinfo{note}{corrected version of
  \textbf{96}(8), 4614--4631}.

\bibitem[{\citenamefont{Bessarab et~al.}(2006)\citenamefont{Bessarab,
  Martynenko, Prudkoi, Soldatov, and Terekhin}}]{BessarabAV:Expser}
\bibinfo{author}{\bibfnamefont{A.~V.} \bibnamefont{Bessarab}},
  \bibinfo{author}{\bibfnamefont{S.~P.} \bibnamefont{Martynenko}},
  \bibinfo{author}{\bibfnamefont{N.~A.} \bibnamefont{Prudkoi}},
  \bibinfo{author}{\bibfnamefont{A.~V.} \bibnamefont{Soldatov}},
  \bibnamefont{and} \bibinfo{author}{\bibfnamefont{V.~A.}
  \bibnamefont{Terekhin}}, \bibinfo{journal}{Radiation Physics and Chemistry}
  \textbf{\bibinfo{volume}{75}}, \bibinfo{pages}{825} (\bibinfo{year}{2006}),
  ISSN \bibinfo{issn}{0969-806X}.

\bibitem[{\citenamefont{Bolotovskii and Serov}(2006)}]{BolotovskiiBM:Radssv}
\bibinfo{author}{\bibfnamefont{B.~M.} \bibnamefont{Bolotovskii}}
  \bibnamefont{and} \bibinfo{author}{\bibfnamefont{A.~V.} \bibnamefont{Serov}},
  \bibinfo{journal}{Radiation Physics and Chemistry}
  \textbf{\bibinfo{volume}{75}}, \bibinfo{pages}{813} (\bibinfo{year}{2006}),
  ISSN \bibinfo{issn}{0969-806X}.

\bibitem[{\citenamefont{Bolotovskii and Ginzburg}(1972)}]{BolotovskiiBM:VaveaD}
\bibinfo{author}{\bibfnamefont{B.~M.} \bibnamefont{Bolotovskii}}
  \bibnamefont{and} \bibinfo{author}{\bibfnamefont{V.~L.}
  \bibnamefont{Ginzburg}}, \bibinfo{journal}{Sov. Phys. Usp.}
  \textbf{\bibinfo{volume}{15}}, \bibinfo{pages}{184} (\bibinfo{year}{1972}).

\bibitem[{\citenamefont{Ginzburg}(1972)}]{GinzburgVL:vaveaa}
\bibinfo{author}{\bibfnamefont{V.~L.} \bibnamefont{Ginzburg}},
  \bibinfo{journal}{Sov. Phys. JETP} \textbf{\bibinfo{volume}{35}},
  \bibinfo{pages}{92} (\bibinfo{year}{1972}), ISSN \bibinfo{issn}{0038-5646}.

\bibitem[{\citenamefont{Bolotovskii and Bykov}(1990)}]{BolotovskiiBM:Radbcm}
\bibinfo{author}{\bibfnamefont{B.~M.} \bibnamefont{Bolotovskii}}
  \bibnamefont{and} \bibinfo{author}{\bibfnamefont{V.~P.} \bibnamefont{Bykov}},
  \bibinfo{journal}{Sov. Phys. Usp.} \textbf{\bibinfo{volume}{33}},
  \bibinfo{pages}{477} (\bibinfo{year}{1990}), ISSN \bibinfo{issn}{0038-5670}.

\bibitem[{\citenamefont{Ardavan}(1998)}]{ArdavanH:Genfnd}
\bibinfo{author}{\bibfnamefont{H.}~\bibnamefont{Ardavan}},
  \bibinfo{journal}{Phys. Rev. E} \textbf{\bibinfo{volume}{58}},
  \bibinfo{pages}{6659} (\bibinfo{year}{1998}), ISSN \bibinfo{issn}{1063-651X}.

\bibitem[{\citenamefont{Ardavan
  et~al.}(2004{\natexlab{b}})\citenamefont{Ardavan, Ardavan, and
  Singleton}}]{ArdavanH:Speapc}
\bibinfo{author}{\bibfnamefont{H.}~\bibnamefont{Ardavan}},
  \bibinfo{author}{\bibfnamefont{A.}~\bibnamefont{Ardavan}}, \bibnamefont{and}
  \bibinfo{author}{\bibfnamefont{J.}~\bibnamefont{Singleton}},
  \bibinfo{journal}{J. Opt. Soc. Am. A} \textbf{\bibinfo{volume}{21}},
  \bibinfo{pages}{858} (\bibinfo{year}{2004}{\natexlab{b}}), ISSN
  \bibinfo{issn}{1084-7529}.

\bibitem[{\citenamefont{Ardavan et~al.}(2007)\citenamefont{Ardavan, Ardavan,
  Singleton, Fasel, and Schmidt}}]{ArdavanH:Morph}
\bibinfo{author}{\bibfnamefont{H.}~\bibnamefont{Ardavan}},
  \bibinfo{author}{\bibfnamefont{A.}~\bibnamefont{Ardavan}},
  \bibinfo{author}{\bibfnamefont{J.}~\bibnamefont{Singleton}},
  \bibinfo{author}{\bibfnamefont{J.}~\bibnamefont{Fasel}}, \bibnamefont{and}
  \bibinfo{author}{\bibfnamefont{A.}~\bibnamefont{Schmidt}},
  \bibinfo{journal}{J. Opt. Soc. Am. A} \textbf{\bibinfo{volume}{24}},
  \bibinfo{pages}{2443} (\bibinfo{year}{2007}).

\bibitem[{not({\natexlab{a}})}]{note:incoherent}
\bibinfo{note}{The superposition of the subbeams is necessarily incoherent
  because the subbeams that are detected at two neighboring points within the
  overall beam arise from two distinct filamentary parts of the source with
  essentially no common elements. The incoherence of this superposition would
  ensure that, though the field amplitude within a subbeam, which narrows with
  distance, decays nonspherically, the field amplitude associated with the
  overall radiation beam, which occupies a constant solid angle, does not.}

\bibitem[{\citenamefont{Jackson}(1999)}]{JacksonJD:Classical}
\bibinfo{author}{\bibfnamefont{J.~D.} \bibnamefont{Jackson}},
  \emph{\bibinfo{title}{Classical Electrodynamics}}
  (\bibinfo{publisher}{Wiley}, \bibinfo{address}{New York},
  \bibinfo{year}{1999}), \bibinfo{edition}{3rd} ed.

\bibitem[{\citenamefont{Hannay}(1996)}]{HannayJH:Bouffr}
\bibinfo{author}{\bibfnamefont{J.~H.} \bibnamefont{Hannay}},
  \bibinfo{journal}{Proc. Roy. Soc. A} \textbf{\bibinfo{volume}{452}},
  \bibinfo{pages}{2351} (\bibinfo{year}{1996}), ISSN \bibinfo{issn}{1364-5021}.

\bibitem[{\citenamefont{Hannay}(2000)}]{HannayJH:ComIGf}
\bibinfo{author}{\bibfnamefont{J.~H.} \bibnamefont{Hannay}},
  \bibinfo{journal}{Phys. Rev. E} \textbf{\bibinfo{volume}{62}},
  \bibinfo{pages}{3008} (\bibinfo{year}{2000}), ISSN \bibinfo{issn}{1063-651X}.

\bibitem[{\citenamefont{Hannay}(2001)}]{HannayJH:ComMhd}
\bibinfo{author}{\bibfnamefont{J.~H.} \bibnamefont{Hannay}},
  \bibinfo{journal}{J. Math. Phys.} \textbf{\bibinfo{volume}{42}},
  \bibinfo{pages}{3973} (\bibinfo{year}{2001}), ISSN \bibinfo{issn}{0022-2488}.

\bibitem[{\citenamefont{Hannay}(2006)}]{HannayJH:Speapc}
\bibinfo{author}{\bibfnamefont{J.~H.} \bibnamefont{Hannay}},
  \bibinfo{journal}{J. Opt. Soc. Am. A} \textbf{\bibinfo{volume}{23}},
  \bibinfo{pages}{1530} (\bibinfo{year}{2006}), ISSN \bibinfo{issn}{1084-7529}.

\bibitem[{\citenamefont{Ardavan et~al.}(2006)\citenamefont{Ardavan, Ardavan,
  and Singleton}}]{ArdavanH:Speapc1}
\bibinfo{author}{\bibfnamefont{H.}~\bibnamefont{Ardavan}},
  \bibinfo{author}{\bibfnamefont{A.}~\bibnamefont{Ardavan}}, \bibnamefont{and}
  \bibinfo{author}{\bibfnamefont{J.}~\bibnamefont{Singleton}},
  \bibinfo{journal}{J. Opt. Soc. Am. A} \textbf{\bibinfo{volume}{23}},
  \bibinfo{pages}{1535} (\bibinfo{year}{2006}), ISSN \bibinfo{issn}{1084-7529}.

\bibitem[{\citenamefont{Morse and Feshbach}(1953)}]{MorsePM:Methods1}
\bibinfo{author}{\bibfnamefont{P.~M.} \bibnamefont{Morse}} \bibnamefont{and}
  \bibinfo{author}{\bibfnamefont{H.}~\bibnamefont{Feshbach}},
  \emph{\bibinfo{title}{Methods of Theoretical Physics}},
  vol.~\bibinfo{volume}{1} (\bibinfo{publisher}{McGraw-Hill},
  \bibinfo{address}{New York}, \bibinfo{year}{1953}).

\bibitem[{\citenamefont{Hoskins}(1999)}]{HoskinsRF:DeltaFn7}
\bibinfo{author}{\bibfnamefont{R.~F.} \bibnamefont{Hoskins}},
  \emph{\bibinfo{title}{Delta Functions: An Introduction to Generalised
  Functions}} (\bibinfo{publisher}{Horwood}, \bibinfo{address}{Chichester},
  \bibinfo{year}{1999}), chap.~\bibinfo{chapter}{7}.

\bibitem[{\citenamefont{Chester et~al.}(1957)\citenamefont{Chester, Friedman,
  and Ursell}}]{ChesterC:Extstd}
\bibinfo{author}{\bibfnamefont{C.}~\bibnamefont{Chester}},
  \bibinfo{author}{\bibfnamefont{B.}~\bibnamefont{Friedman}}, \bibnamefont{and}
  \bibinfo{author}{\bibfnamefont{F.}~\bibnamefont{Ursell}},
  \bibinfo{journal}{Proc. Camb. Phil. Soc.} \textbf{\bibinfo{volume}{53}},
  \bibinfo{pages}{599} (\bibinfo{year}{1957}).

\bibitem[{\citenamefont{Burridge}(1995)}]{BurridgeR:Asyeir}
\bibinfo{author}{\bibfnamefont{R.}~\bibnamefont{Burridge}},
  \bibinfo{journal}{SIAM J. Appl. Math.} \textbf{\bibinfo{volume}{55}},
  \bibinfo{pages}{390} (\bibinfo{year}{1995}), ISSN \bibinfo{issn}{0036-1399}.

\bibitem[{\citenamefont{Bleistein and Handelsman}(1986)}]{BleisteinN:Asei}
\bibinfo{author}{\bibfnamefont{N.}~\bibnamefont{Bleistein}} \bibnamefont{and}
  \bibinfo{author}{\bibfnamefont{R.~A.} \bibnamefont{Handelsman}},
  \emph{\bibinfo{title}{Asymptotic Expansions of Integrals}}
  (\bibinfo{publisher}{Dover}, \bibinfo{address}{New York},
  \bibinfo{year}{1986}).

\bibitem[{not({\natexlab{b}})}]{note:gradcomp}
\bibinfo{note}{That these components of the gradient are of the same order of
  magnitude is a consequence of the fact that the spiraling cusps that emanate
  from this source propagate to infinity along a conical surface centered at
  the origin and so have nonzero pitch angles.}

\bibitem[{\citenamefont{Popov et~al.}(2006)\citenamefont{Popov, Soglasnov,
  Kondrat'ev, Kostyuk, and Ilyasov}}]{PopovMV:GPmcrecp}
\bibinfo{author}{\bibfnamefont{M.~V.} \bibnamefont{Popov}},
  \bibinfo{author}{\bibfnamefont{V.~A.} \bibnamefont{Soglasnov}},
  \bibinfo{author}{\bibfnamefont{V.~I.} \bibnamefont{Kondrat'ev}},
  \bibinfo{author}{\bibfnamefont{S.~V.} \bibnamefont{Kostyuk}},
  \bibnamefont{and} \bibinfo{author}{\bibfnamefont{Y.~P.}
  \bibnamefont{Ilyasov}}, \bibinfo{journal}{Astron. Rep.}
  \textbf{\bibinfo{volume}{50}}, \bibinfo{pages}{55} (\bibinfo{year}{2006}),
  ISSN \bibinfo{issn}{1063-7729}.

\bibitem[{\citenamefont{Hankins et~al.}(2003)\citenamefont{Hankins, Kern,
  Weatherall, and Eilek}}]{HankinsTH:Nanrbs}
\bibinfo{author}{\bibfnamefont{T.~H.} \bibnamefont{Hankins}},
  \bibinfo{author}{\bibfnamefont{J.~S.} \bibnamefont{Kern}},
  \bibinfo{author}{\bibfnamefont{J.~C.} \bibnamefont{Weatherall}},
  \bibnamefont{and} \bibinfo{author}{\bibfnamefont{J.~A.} \bibnamefont{Eilek}},
  \bibinfo{journal}{Nature (London)} \textbf{\bibinfo{volume}{422}},
  \bibinfo{pages}{141} (\bibinfo{year}{2003}), ISSN \bibinfo{issn}{0028-0836}.

\bibitem[{\citenamefont{Soglasnov et~al.}(2004)\citenamefont{Soglasnov, Popov,
  Bartel, Cannon, Novikov, Kondratiev, and Altunin}}]{SoglasnovVA:GiapPB}
\bibinfo{author}{\bibfnamefont{V.~A.} \bibnamefont{Soglasnov}},
  \bibinfo{author}{\bibfnamefont{M.~V.} \bibnamefont{Popov}},
  \bibinfo{author}{\bibfnamefont{N.}~\bibnamefont{Bartel}},
  \bibinfo{author}{\bibfnamefont{W.}~\bibnamefont{Cannon}},
  \bibinfo{author}{\bibfnamefont{A.~Y.} \bibnamefont{Novikov}},
  \bibinfo{author}{\bibfnamefont{V.~I.} \bibnamefont{Kondratiev}},
  \bibnamefont{and} \bibinfo{author}{\bibfnamefont{V.~I.}
  \bibnamefont{Altunin}}, \bibinfo{journal}{Astrophys. J.}
  \textbf{\bibinfo{volume}{616}}, \bibinfo{pages}{439} (\bibinfo{year}{2004}),
  ISSN \bibinfo{issn}{0004-637X}.

\bibitem[{\citenamefont{Schmidt et~al.}(2007)\citenamefont{Schmidt, Ardavan,
  Fasel, Singleton, and Ardavan}}]{SchmidtA:Occopmlw}
\bibinfo{author}{\bibfnamefont{A.}~\bibnamefont{Schmidt}},
  \bibinfo{author}{\bibfnamefont{H.}~\bibnamefont{Ardavan}},
  \bibinfo{author}{\bibfnamefont{J.}~\bibnamefont{Fasel}},
  \bibinfo{author}{\bibfnamefont{J.}~\bibnamefont{Singleton}},
  \bibnamefont{and} \bibinfo{author}{\bibfnamefont{A.}~\bibnamefont{Ardavan}},
  in \emph{\bibinfo{booktitle}{Proceedings of the 363rd WE-Heraeus Seminar on
  Neutron Stars and Pulsars}}, edited by
  \bibinfo{editor}{\bibfnamefont{W.}~\bibnamefont{Becker}} \bibnamefont{and}
  \bibinfo{editor}{\bibfnamefont{H.~H.} \bibnamefont{Huang}}
  (\bibinfo{year}{2007}), pp. \bibinfo{pages}{124--127},
  \eprint{astro-ph/0701257}.

\end{thebibliography}

\end{document}